\documentclass[oneside,english]{amsart}
\usepackage[T1]{fontenc}
\usepackage[latin9]{inputenc}
\usepackage{url}
\usepackage{float}
\usepackage{texnansi}
\usepackage{color}
\usepackage{tikz}
\usepackage{subfigure}
\usepackage{enumerate}
\usepackage{algorithm}
\usepackage{algorithmic}
\usepackage[normalem]{ulem}
\usepackage{graphicx}
\usepackage{lmodern}
\usepackage{amsmath}
\usepackage{amssymb}
\usepackage{amsfonts}
\usepackage{graphicx}
\usepackage{setspace}
\usepackage{mathrsfs}
\usepackage{multirow}
\usepackage{dsfont}
\usepackage{array}
\usepackage{amsthm}

\makeatletter
\numberwithin{equation}{section}
\numberwithin{figure}{section}
\theoremstyle{plain}

  \theoremstyle{plain}

\usepackage{dsfont}
\usepackage{mathrsfs}

\newcommand{\bOne}{{\mathbf 1}}
\newcommand{\Pb}{{\mathbb P}}
\newcommand{\E}{{\mathbb E}}
\newcommand{\lambdahat}{\hat{\lambda}}
\newcommand{\normtwo}[1]{\lVert #1 \rVert_2}

\makeatother

\usepackage{babel}

\begin{document}

\title{Sparse choice models}\thanks{This work is supprted in part by NSF CMMI project. This
work was done while all authors were at the Massachusetts Institute of Technology (MIT). VF is with the Sloan School of Management, MIT; DS is with department of EECS, MIT; and SJ is 
with the Stern School of Business, NYU}

\author{Vivek F. Farias \ \ Srikanth Jagabathula \ \  Devavrat Shah}

\maketitle
\global\long\def\norm#1{\left\Vert #1\right\Vert }

\global\long\def\normzero#1{\left\Vert #1\right\Vert _{0}}

\global\long\def\beps{\varepsilon}

\global\long\def\supp#1{\mathsf{supp}\left(#1\right)}

\global\long\def\crF{\mathscr{F}}

\global\long\def\Ascr{\mathcal{A}}

\global\long\def\abs#1{\left|#1\right|}

\global\long\def\defas{\overset{\mathrm{def}}{=}}

\global\long\def\crE{\mathscr{E}}

\global\long\def\Fscr{\mathcal{F}}

\global\long\def\crS{\mathscr{S}}

\global\long\def\Var{\mathrm{Var}}

\global\long\def\Mscr{\mathcal{M}}

%theorem definitions
\newtheorem{theorem}{Theorem}[section]
\newtheorem{lemma}{Lemma}[section]
\newtheorem{definition}{Definition}[section]
\newtheorem{example}{Example}[subsection]
\newtheorem{property}{Property}
\newtheorem{hypothesis}{Hypothesis}
\newtheorem{condition}{Condition}
\newtheorem{claim}{Claim}
\newenvironment{random}[2][Random Model]{\begin{trivlist}
\item[\hskip \labelsep {\bfseries #1}\hskip \labelsep {\bfseries #2}]}{\end{trivlist}}

%local definitions
\newcommand{\sparsity}{K}
\newcommand{\set}[1]{\left\{ #1 \right\}}
\newcommand{\rep}{M^{\lambda}}
\newcommand{\oracle}{\mathsf{oracle}}
\newcommand{\true}{\operatorname{\mathsf{true}}}
\newcommand{\false}{\operatorname{\mathsf{false}}}
\newcommand{\feasperms}{\operatorname{\mathsf{feas-perms}}}
\newcommand{\mustperms}{\operatorname{\mathsf{must-perms}}}
\newcommand{\candperms}{\operatorname{\mathsf{cand-perms}}}
\newcommand{\bestset}{\operatorname{\mathsf{best-set}}}
\newcommand{\besterror}{\operatorname{\mathsf{best-error}}}
\newcommand{\Kscr}{\mathcal{K}}
\newcommand{\algo}{\text{{\sf ALGO}}}

\begin{abstract}

Choice models, which capture popular preferences over objects of
interest, play a key role in making decisions whose eventual outcome is impacted by human choice behavior. In most scenarios, 
the choice model, which can effectively be viewed as a distribution over 
permutations, must be learned from observed data. The observed data, in turn, may frequently be viewed as (partial, noisy) information about marginals of this distribution over permutations. 
As such, the search for an appropriate choice model boils down to learning a distribution over 
permutations that is (near-)consistent with observed information about this distribution. 

In this work, we pursue a non-parametric approach which seeks to
learn a choice model (i.e. a distribution over permutations) with {\em sparsest} possible support, and consistent with
observed data. We assume that the data observed consists of noisy information pertaining to the marginals of the choice model we seek to learn. 
We establish that {\em any} choice model admits a `very' sparse approximation in the sense that there exists a choice model whose support is small relative to the dimension of the observed data and whose marginals approximately agree with the observed marginal information. 
We further show that under, what we dub, `signature' conditions, such a sparse approximation can be found in a computationally efficiently fashion relative to a brute force approach. 
An empirical study using the American Psychological
Association election data-set suggests that our approach manages to unearth useful 
structural properties of the underlying choice model using the sparse approximation found. 
Our results further suggest that the signature condition is a potential alternative to the 
recently popularized Restricted Null Space condition for efficient
recovery of sparse models.

\end{abstract}

\section{Introduction}

\subsection{Background}

It is imperative for an architect of a societal system, be it a road
transportation system, energy distribution network, or the Internet, to deal
with the uncertainty arising from human participation in general, and human choice behavior in particular. One possible approach
to serve this end, is to make assumptions on the behavior of an individual (for instance, assuming that every individual is a rational utility maximizer). Such an assumption leads, in
turn, to a collective behavioral model for the entire population. This
model can subsequently be used to guide system design, e.g. where to
invest resources to build new roads in the country or what sorts of products to put up for sale at a store. Such models, of the collectives preference of a population over objects of interest, are colloquially referred to as customer choice models, or simply choice models. As suggested by the above discussion, choice models form crucial inputs to making effective decisions across a swathe of domains. 
%In all such decisions, effectively, what
%matters is the collective preferences of the population over objects of
%interest, which are captured by the so called customer choice model. 
%Therefore,
%the collective behavioral model induces a choice model, which in turn makes it
%possible to make such decisions effectively. 
%by knowing this choice model, it is possible to make such decision effectively.

Now in practice, a choice model is revealed through partial information people provide about their preferences via their purchase behavior, responses to polls, or in explicit choices they make. 
%partial preferences
%that people express either through purchase behavior, polled opinions or
%explicitly revealed judgement. 
In assuming a behavioral model
for the population one runs the risk of mis-modeling choice. Ideally, one wishes to learn a choice model consistent with observed partial preferences, having made little or no behavioral or structural assumptions. In the absence of such structural assumptions one needs a criterion to select a choice model from among the many that will likely agree with the observed partial preferences. A natural criterion here is structural `simplicity' (a precise definition of which we defer for now). Since choice models are used as inputs to decision problems, it makes operational sense to seek a choice model that is structurally simple. In addition, a criterion of this sort is consistent with Occam's razor.
%and running the risk of being terribly poor in prediction of society, 
%a more concrete way to approach the aforementioned decision problems is to first
%{\em learn} choice model from the {\em observed} partial preferences and then
%use it to make decisions. The main challenge in such an approach is to learn a
%choice model that is consistent with partial preferences. 
% Furthermore, the search for
%a simple model to explain the observed partial preferences is consistent with
%the Occam's razor philosophy. 
Thus motivated, we consider here the question of efficiently
learning a `simple' choice model that is consistent with observed partial
(marginal) information.
% refore, effectively the challenge in such an approach bolis down to learning a
% choice model consistent with partial preferences. 
% Now the choice model is usually used as input for decision making. Therefore,
% for computational efficiency reason, it makes sense to seek out a choice model
% that is structurally simple. Further search for simple choice model explaining
% observation re-inforces the philosophy of Occam's razor. For all these reasons,
% we consider the question of efficiently learning sparse choice model consistent
% with observed partial (marginal) information.
\subsection{Related prior work}
There is a large literature devoted to learning structurally simple choice models from 
partial observations. Most prior work has focused on parametric
approaches. 
% The primary focus of the prior works has been variety of
% parametric choice models. 
Given the nature of the topic, the literature is quite diverse and hence it is
not surprising that the same choice model appears under different names in
different areas. In what follows, we provide a succinct overview of the
literature.

\subsubsection{Learning Parametric Models}

To being with, the monograph by Diaconis \cite[Chapter 9]{diaconis-book}
provides a detailed history of most of the models and references given below. In
the simplest setting, a choice model (which, recall, is simply a distribution over the permutations
of $N$ objects of interest) is captured by the order statistics of $N$ random
variables $Y_1,\dots, Y_N$. Here $Y_i = u_i + X_i$ where the $u_i$ are
parameters and the $X_i$ are independent, identically distributed random
variables. Once the distributions of the $X_i$s are specified, the choice model is
specified.

This class of models was proposed nearly a century ago by
Thurstone~\cite{thurstone}. A specialization of the above model when the $X_i$s
are assumed to be normal with mean $0$ and variance $1$ is known as the
Thurstone-Mosteller model. This is also known, more colloquially, as the {\em probit} model.

Another specialization of the Thurstone model is realized when the $X_i$s are
assumed to have Gumbel or Logit distributions (one of the extreme value
distribution). This model is attributed differently across communities.
Holman and Marley established that this model is equivalent (see \cite{yellott}
for details) to a generative model where the $N$ objects have positive weights
$w_1,\dots, w_N$ associated with them, and a random permutation of these $N$
objects is generated by recursive selection (without replacement) of objects in
the first position, second position and so on with selection probabilities
proportional to theirs weights. As per this, the probability of object $i$ being
preferred over object $j$ (i.e. object $i$ is ranked higher compared to object
$j$) is $w_i/(w_i+w_j)$. The model in this form is known as the Luce model
\cite{luce} as also the Plackett model \cite{plackett}\footnote{It is worth noting that this model is very similar the the Bradley-Terry model \cite{BR}, where each object $i$ has weight $w_i > 0$ associated with
  it; the Bradley-Terry model is however distinct from the model proposed by Plackett and
  Luce in the probabilities it assigns to each of the permutations.}. Finally, this model is also refereed to as the Multinomial Logit Model (MNL) after McFadden referred to it as a {\em
  conditional logit} model \cite{McF1}; also see \cite{Debreu60}. We will adopt the convention of referring to this important model as the MNL model. 

The MNL model is of central importance for
various reasons. It was introduced by Luce to be consistent with the axiom of
{\em independence from irrelevant alternatives} (IIA). The model was shown to be
consistent with the induced preferences assuming a random utility
maximization framework whose inquiry was started by Marschak \cite{Marschak1,
  Marschak}. Very early on, simple statistical tests as well as simple
estimation procedure were developed to fit such a model to
observed data \cite{McF1}. % 65 paper
Now the IIA property possessed by the MNL model is not necessarily desirable as evidenced by empirical studies. Despite such structural limitations, the MNL model has been {\em widely} utilized across application areas primarily due to the
ability to learn the model parameters easily from observed data. For example,
see \cite{M81, BL85, McF2} for application in transportation and \cite{GL83,
  Mahajan99} for applications in operations management and marketing.

With a view to addressing the structural limitations of the MNL model, a number of generalizations to this model have been proposed over the years. Notable among these is the so-called `nested' MNL model, as well as mixtures of MNL models (or MMNL models). These generalizations avoid the IIA property and continue to be consistent with the random utility maximization framework at the expense of increased model complexity; see \cite{B73, BL85, BM80, CD80, MT00}.  The interested reader is also referred to an overview article on this line of research by McFadden \cite{McF2}. 
% led to further refinements to yield models
%such as the nested version of the MNL model, or mixture of MNL models. These
%models don't exhibit the IIA property and also fit the broad random utility
%maximization framework thus providing consistency of empirical as well as
%behavioral view of population, see \cite{B73, BL85, BM80, CD80, MT00} for
%example. An interested reader is referred to an overview article on this line of
%development by McFadden \cite{McF2}. 
While generalized models of this sort are in
principle attractive, their complexity makes them difficult to learn while avoiding the risk of over-fitting. 
%they are either too complicated and hence difficult to
%learn or they impose too much undesirable structure. 
More generally,
specifying an appropriate parametric model is a difficult task, and the risks
associated with mis-specification are costly in practice. For an applied view of
these issues see \cite{Bartels99, Horowitz93, Debreu60}. Thus, while these
models are potentially valuable in specific well understood scenarios, the
generality of their applicability is questionable.

As an alternative to the MNL model (and its extensions), one might also consider the parametric family of choice models induced by the exponential family of distributions over permutations. 
%We take note of parametric models induced by the exponential family of models.
These may be viewed as choice models that have maximum entropy among those models that satisfy the constraints imposed by the observed data. The number of parameters in such a model is equal to the number of constraints in the maximum entropy optimization formulation, or equivalently the effective dimension of the
underlying data (cf. the Koopman-Pitman-Darmois Theorem~\cite{KPD}).
%MNL model can be viewed as the simplest possible member of this family. 
This scaling of the number of parameters with the effective data
dimension makes the exponential family obtained via the maximum entropy principle
very attractive. 
% property of the number of parameters scaling proportional to the effective
% data dimension makes the exponential family obtained via maximum entropy
% principle very attractive. 
Philosophically, this approach imposes on the choice model, only 
those constraints implied by the observed data. On the flip side, learning the
parameters of an exponential family model is a computationally challenging task
(see \cite{Crain}, \cite{Beran} and \cite{JW08}) as it requires computing a
``partition function'' possibly over a complex state space.

\subsubsection{Learning Nonparametric Models}

As summarized above, parametric models either impose strong restrictions on the structure
of the choice model and/or are computationally challenging to learn. To overcome
these limitations, we consider a nonparametric approach to learning a choice
model from the observed partial data.
%or distribution over permutations from observed partial or limited data.

The given partial data most certainly does not completely identify the
underlying choice model or distribution over permutations. Specifically, there
are potentially multiple choice models that are (near) consistent with the given
observations, and we need an appropriate model selection criterion. In the
parametric approach, one uses the imposed parametric structure as the model
selection criterion. On the other hand, in the nonparametric approach considered
in this paper, we use {\em simplicity}, or more precisely the {\em sparsity} or support size of the distribution over permutations, as the criterion for selection:
specifically, we select the {\em sparsest} model (i.e. the distribution with the
smallest support) from the set of models that are (near) consistent with the
observations. This nonparametric approach was first proposed by Jagabathula and
Shah~\cite{JS08, JS2} and developed further by Farias, Jagabathula and
Shah~\cite{FJS1, MSpaper}. Following~\cite{JS08, JS2, FJS1, MSpaper}, we
restrict ourselves to observations that are in the form of marginal information
about the underlying choice model. For instance, the observations could be in
the form of {\em first-order} marginal information, which corresponds to
information about the fraction of the population that ranks object $i$ at position
$j$ for all $1 \leq i,j \leq N$, where $N$ is the number of objects.

% Now given observation about partial information regarding the choice model, there 
% are likely to be multiple choice models that are (near) consistent with it. Therefore, 
% it is necessary to choose one of these many possible choice models. A parametric 
% approach makes such selection by imposing its parametric structure on the 
% observations. In the non-parametric approach considered in this paper, we use 
% {\em simplicity} or more precisely {\em sparsity} (small support of the distribution) 
% as the criteria for such selection: the distribution over permutations that is consistent
% with the observation and has the smallest possible support. This non-parametric 
% approach was first proposed by Jagabathula and Shah \cite{JS08, JS2} and 
% developed further by Farias, Jagabathula and Shah \cite{FJS1, MSpaper}. 
% Following \cite{JS08, JS2, FJS1, MSpaper}, in this paper we shall restrict ourselves
% to the scenarios where the observations made about choice model are in the form of 
% the marginal distributions, e.g. the {\em first-order} marginal distribution of
% a choice model over $N$ objects provides information about what fraction 
% of population ranks object  $i$ in the $j^{th}$ position for all $1\leq i, j \leq N$ ?

A major issue with the identification of sparse models from marginal information
is the associated computational cost. Specifically, recovering a distribution
over permutations of $N$ objects, in principle, requires identifying
probabilities of $N!$ distinct permutations. The distribution needs to be
recovered from marginal information, which can usually be cast as a lower
dimensional ``linear projection'' of the underlying choice model; for instance,
the first-order marginal information can be thought of as a linear projection of
the choice model on the $(N-1)^2$ dimensional space of doubly stochastic matrices. Thus, finding a sparse
model consistent with the observations is equivalent to solving a severely
underdetermined system of linear equations in $N!$ variables, with the aim of
finding a sparse solution. As a result, at a first glance, it appears that the
computational complexity of {\em any} procedure should scale with the dimension
of the variable space, $N!$. %[TODO: KEEP THIS SENTENCE?]

In \cite{JS08, JS2, FJS1, MSpaper}, the authors identified a so called
`signature condition' on the space of choice models and showed that whenever a
choice model satisfies the signature condition and {\em noiseless} marginal data
is available, it can be {\em exactly} recovered in an efficient manner from
marginal data with computational cost that scales linearly in the dimension of the
marginal data ($(N-1)^2$ for first-order marginals) and exponentially in the
sparsity of the choice model. Indeed, for sparse choice models this is
excellent. They also established that the `signature condition' is not merely a
theoretical construct. In fact, a randomly chosen choice model with a
``reasonably large'' sparsity (support size) satisfies the `signature condition'
with a high probability. The precise sparsity scaling depends on the type of
marginal data available; for instance, for the first-order marginals, the
authors show that a randomly generated choice model with sparsity up to $O(N\log
N)$ satisfies the `signature conditions'. In summary, the works of
\cite{JS08,JS2,FJS1, MSpaper} establish that if the original choice model
satisfies the `signature condition' (e.g. generated randomly with reasonable
sparsity) and the available observations are {\em noise-free}, then the sparsest
choice model consistent with the observations can be recovered efficiently.

\subsection{Our contributions}

In reality, available data is not noise-free. Even more importantly, the data might
arise from a distribution that is {\em not} sparse to begin with! The main
contribution of the present paper is to address the problem of learning non-parametric choice models in the challenging, more realistic case when the underlying model is potentially non-sparse and the marginal data is corrupted by noise.
%extend the prior works~\cite{JS08,JS2,FJS1,
%  MSpaper} to the more realistic case when the underlying model may not be
%sparse and the marginal data may be corrupted by noise. 
Specifically, we
consider the problem of finding the sparsest model that is near consistent with
-- or equivalently, within a ``distance'' $\beps$ of -- the marginal data. We
consider the setting in which the marginal information can be cast as a linear
projection of the underlying choice model over a lower dimensional space. We
restrict ourselves primarily to {\em first-order} marginal information
throughout this paper; a discussion about how our methods and results extend to
general types of marginal information is deferred to the end.

In this context, we consider two main questions: (1) How does one find the
sparsest consistent distribution in an efficient manner? and (2) How ``good''
are sparse models in practice? Next, we describe the contributions we make towards
answering these questions.

In order to understand how to efficiently find sparse models approximating the
given data, we start with the more fundamental question of ``how sparse can the
sparsest solution be?'' To elaborate further, we first note that the space of
first-order marginal information is equivalent to the space of doubly stochastic
matrices (this equivalence is explained in detail in subsequent sections). Given
this, finding the sparsest choice model that is near consistent with the observations
is essentially equivalent to determining the convex decompositions (in terms of permutations) of all doubly
stochastic matrices that are within a ball of ``small'' radius around the given
observation matrix and choosing the sparsest convex decomposition. Now, it
follows from the celebrated Birkhoff-von Neumann's theorem (see~\cite{BrK} and
\cite{VN}) that a doubly stochastic matrix belongs to an $(N-1)^2$ dimensional
polytope with the permutation matrices as the extreme points. Therefore
Caratheodory's theorem tells us that it is possible to find a convex
decomposition of any doubly stochastic matrix with at most $(N-1)^2 + 1$ extreme
points, which in turn implies that the sparsest model consistent with the
observations has a support of at most $(N-1)^2 + 1 = \Theta(N^2)$. We raise the
following natural question at this point: given {\em any} doubly stochastic
matrix (i.e. any first-order marginal data), does there exist a near
consistent choice model with sparsity significantly smaller than $\Theta(N^2)$?

%Note that the answer to this question can give us an indication of whether a
%straightforward approach (such as convex relaxation) can produce good
%approximations to the sparsest solution. In particular, it could be possible
%that for a general doubly stochastic matrix, there does not exist near
%consistent models with sparsity significantly smaller than $\Theta(N^2)$. If
%such is the case, then convex relaxations -- which could result in a models of
%sparsity $O(N^2)$ -- can produce solutions close to the optimal, at least for
%general observation matrices. In that case, we could attempt to characterize the
%class of observation matrices for which the sparsest model can be identified
%through convex relaxation. If, on the other hand, we can prove that for a
%general observation matrix, the sparsest model can have sparsity significantly
%smaller than $\Theta(N^2)$, then using straightforward approaches like convex
%relaxations can result in a highly suboptimal solutions. 

Somewhat surprisingly, we establish that in as much as first-order marginals are concerned, {\em any} choice model can be $\beps$-approximated by a choice model with sparsity or support $O(N/\beps^2)$. More specifically, we show that any non-negative valued doubly stochastic matrix can be $\beps$-approximated in the $\ell_2$ sense by a convex combination of $O(N/\beps^2)$ permutation matrices. This is significantly smaller than $\Theta(N^2)$.

%(in the sense that the first order marginals of the approximating distribution are 
%the first-order marginal of any
%choice model (equivalently any non-negative valued doubly stochastic matrix)
%can be $\beps$-approximated (with respect to $\ell_2$-approximation) by 
%first-order marginal of a choice model with sparsity or support $O(N/\beps^2)$.

%%%%%%%%%

The next question pertains to finding such a sparse model efficiently. As mentioned
above, the signature conditions have played an important role in efficient learning
of sparse choice models from noise-free observations. It is natural to ask whether
they can be useful in the noisy setting. More precisely, can the first-order
marginals observed be well approximated by those of a choice model from the signature
family, and if so, can this be leveraged to efficiently recover a sparse choice model consistent with observations. 
%utilized to advantage for computationally efficient 
%recovery of a sparse choice model consistent with observations. 

To answer the first question,  we identify conditions on the original choice models such that they admit $\beps$-approximations by sparse choice models that satisfy the `signature' conditions and 
that have sparsity $O(N/\beps^2)$. We establish that for a very 
large class of choice models, including very dense models such as the models from the MNL 
or exponential family, the observed marginal information can be well approximated by a sparse  
choice model from the signature family. We are then able to use this result to our advantage in designing a novel algorithm for the recovery of a sparse choice model given noisy first order marginal information. In particular, in leveraging this result, our algorithm uses structural properties of models in the signature family along with an adaptation of the 
multiplicative weights update framework of Plotkin-Shmoys-Tardos \cite{PST}. Our algorithm finds a sparse choice model with sparsity $O(K \log N)$ in time 
$\exp\big(\Theta(K \log N)\big)$ if there exists a choice model in the signature family with sparsity $K$ that approximates the data well; our structural result guarantees the existence of such approximations for suitable $K$.  

%To answer the second question, we need to extend the prior works \cite{JS08, JS2, FJS1, MSpaper} 
%by developing computationally efficient recovery of sparse choice model from signature
%family based on noisy first-order marginals. We develop a novel algorithm utilizing the 
%multiplicative weights update framework of Plotkin-Shmoys-Tardos \cite{PST} and the 
%classical result of Birkhoff and Von Neumann \cite{BrK, VN} about permutations being 
%the extreme points of the space of doubly stochastic  matrices. 
%
%This algorithm, 
%effectively finds a sparse choice model with sparsity $O(K \log N)$ in time 
%$\exp\big(\Theta(K \log N)\big)$ if there exists a choice model in signature 
%family with sparsity $K$ that approximates data well. 

To start with, this is much (exponentially in $N$) better than the brute force
search which would require ${N! \choose K}$ $\approx \exp\big(\Theta(KN \log N) \big)$
computation. Given that for a large class of choice models, their marginal data
is well approximated by a signature family choice model with sparsity essentially
$O(N)$, the computation cost is bounded by $\exp\big(O(N \log N)\big)$ which
is $\big(N!\big)^{O(1)}$ -- polynomial in the dimension, $N!$, of the ambient data. 
This is on an equal footing with many of the recently developed sparse model learning
methods under the framework of {\em compressed sensing}. These latter methods are based on 
convex optimization (typically, linear programming) and have computational cost that grows polynomially in the 
dimension of the ambient data.  %[TODO: EDIT?]

We establish the effectiveness of our approach by applying our sparse model 
learning procedure to the well studied American Psychological Association's (APA)
ranked election data (i.e., the data used by Diaconis in \cite{diaconis89}). Interestingly enough,
through  sparse model approximation of the election data, we find structural
information in the data similar to that unearthed by Diaconis \cite{diaconis89}. 
The basic premise in \cite{diaconis89} was that by looking at linear projections
of the ranked election votes, it may be possible to unearth hidden structure
in the data. Our sparse approximation captures similar structural information 
from projected data suggesting the utility of this approach in unearthing non-obvious
structural information.

\subsubsection{Thematic Relation: Compressed Sensing}

We note that this work is thematically related to the recently developed theory
of compressed sensing and streaming algorithms, and further to classical coding
theory and signal processing cf. \cite{S49, N02}. In the compressive sensing
literature (see \cite{CT05, CRT06, CR06, CRT06R, D06}), the goal is to estimate
a `signal' by means of a minimal number of measurements. Operationally this is
equivalent to finding the sparsest signal consistent with the observed (linear)
measurements. In the context of coding theory, this corresponds to finding the
most likely transmitted codeword given the received message~\cite{G62, RS60,
  SS96, LMSS01}. In the context of streaming algorithms, this task corresponds
to maintaining a `minimal' data structure to implement algorithmic
operations~\cite{T06, T04, BGIK08, CM06, GSTV07}. In spite of the thematic
similarity, existing approaches to compressive sensing are ill-suited to the
problem at hand; see~\cite{JS2}, where the authors establish that the generic
{\em Restricted Null Space} condition -- a necessary and sufficient condition
for the success of the convex optimization in finding sparsest possible solution
-- is not useful in the setting considered here. In a nutshell, the `projections' of the signal we observe are a given as opposed to being a design choice. Put another way, the present
paper can be viewed as providing a non-trivial extension to the theory of
compressive sensing for the problem of efficiently learning distributions over
permutations.

\subsection{Organization}

The rest of the paper is organized as follows. In Section~\ref{sec:setup}, the
precise problem statement along with the signature condition is introduced. We
also introduce the MNL and exponential family parametric models. The main
results of this paper are stated in Section~\ref{sec:results}. These results are
established in Section~\ref{sec:proofs}. In Section~\ref{sec:empirical}, we
study an application of our results to the popular benchmark APA data set.
Using a simple heuristic motivated by the signature condition, we learn a sparse
approximation of the observed data. We discuss the relevance of the sparse approximation thus obtained and conclude that it provides positive support to the quest
of searching for sparse choice models. Finally, we conclude in
Section~\ref{sec:discussion} with a discussion on how the methods we propose for
first-order marginal data extend to general types of marginal data.

%
%
%
%
%
%Our approach is to
%formulate the search for the solution as an integer program. Thus,
%in Section~\ref{sub:signature_family} , we describe a representation
%of the signature family. In Section~\ref{sub:noiseless}, we describe
%the integer program we use to find the solution when the partial information
%is generated by a distribution in the signature family. Then, in Section~\ref{sub:general_case},
%we describe the general integer program for finding the sparsest distribution
%in the signature family that well-approximates the given partial information
%vector. Finally, in Section~\ref{sec:guarantees} we prove guarantees
%for our procedure for the case when the partial information vector
%$y$ corresponds to first-order information. 

\section{Setup} \label{sec:setup}

Given $N$ objects or items, we are interested in a choice model or distribution
over permutations of these $N$ items. Let $S_N$ denote the space of $N!$
permutations of these $N$ items. A choice model (equivalently, a distribution
over $S_N$) can then be represented as a vector of $N!$ dimension with
non-negative components, all of them summing up to $1$. The observations we
consider here are certain marginal distributions of the choice model. Specifically,
throughout this paper, we primarily restrict ourselves to {\em first-order}
marginal information. We point out how our results extend to general marginal
information in the discussion (Section~\ref{sec:discussion}).

More precisely, let $\lambda$ denote a choice model or a distribution over
$S_N$. Then, the first-order marginal information, $M(\lambda) = [M_{ij}(\lambda)]$, is
an $N\times N$ doubly stochastic matrix with non-negative entries
defined as 
\[
M_{ij}(\lambda) = \sum_{\sigma \in S_N} \lambda(\sigma) \bOne_{\{\sigma(i) = j\}},
\]
where $\sigma \in S_N$ represents a permutation, $\sigma(i)$ denotes the rank of
item $i$ under permutation $\sigma$, and $\bOne_{\{x\}}$ is the standard
indicator with $\bOne_{\{\mbox{\footnotesize true}\}} = 1$ and
$\bOne_{\{\mbox{\footnotesize false}\}} = 0$.

We assume that there is a {\em ground-truth} choice model $\lambda$ and 
the observations are a noisy version of $M$. Specifically, 
let the observations be $D = M + \eta$ so that $\|\eta\|_2 \leq \delta$ 
for some small enough $\delta > 0$; by $\|\eta\|_2$ we mean
\[
\| \eta\|_2^2 = \sum_{i, j = 1}^N \eta_{ij}^2. 
\]
Without loss of generality, we assume that $D$ is also doubly
stochastic (or else, it is possible to transform it into that form). 
The goal is to learn a choice model or distribution $\hat{\lambda}$ over
$S_N$ so that it's first-order marginal ${M}(\hat{\lambda})$ approximates
$D$ (and hence $M$) well and the support of $\hat{\lambda}$, 
$\|\hat{\lambda}\|_0$ is small. Here 
\[
\|\hat{\lambda}\|_0 \stackrel{\triangle}{=} \big|\{ \sigma \in S_N: \hat{\lambda}(\sigma) > 0 \} \big|.
\]
Indeed, one way to find such a $\hat{\lambda}$ is to solve the following program: for a choice of approximation
error $\beps > 0$, 
\begin{align}
\text{\sf minimize} & \quad \|\mu\|_0 \qquad \text{\sf over} \quad \text{choice~models~}~\mu \label{eq:P0} \\
\text{\sf such~that} & \quad \| M(\mu) - D\|_2 \leq \beps. \nonumber
\end{align}
By the Birkhoff-Von Neumann theorem and Caratheodary's theorem (as discussed
earlier), there must exist a solution, say $\hat{\lambda}$, to \eqref{eq:P0}
with $\|\hat{\lambda}\|_0 \leq (N-1)^2 + 1$. Therefore, a solution to 
program \eqref{eq:P0} with guarantee $\|\hat{\lambda}\|_0 \leq (N-1)^2 + 1$
can be achieved by a linear programing relaxation of \eqref{eq:P0}. The basic 
question is, whether such a solution is near optimal. Putting another way

\medskip
\noindent{\bf Question 1.} Given {\em any} doubly stochastic matrix $D$, does
there exist a choice model $\lambda$ with sparsity significantly smaller than
$\Theta(N^2)$ such that $\normtwo{M(\lambda) - D} \leq \beps$.
%, for any doubly stochastic matrix $M$?
\medskip

Geometrically speaking, the question above translates to: given a ball of radius
$\beps$ around $D$, is there a subspace spanned by $K$ extreme points that
intersects the ball, for {\em any} doubly stochastic matrix $D$ and some $K$
that is significantly smaller than $\Theta(N^2)$? 

\medskip
Now if the sparsest solution has sparsity $K$, then the brute-force approach 
would require searching over $\binom{N!}{K} \approx
\exp(\Theta(KN\log N))$ options. The question here is whether this could
be improved upon significantly. That is, 

\medskip
\noindent {\bf Question 2.} Is it possible to solve~\eqref{eq:P0}
with a running time complexity that is far better than $\exp(\Theta(K N \log N))$, at
least for a reasonable large class of observations $D$?
\medskip

We obtain a faster algorithm by restricting our search to models that belong to
the signature family that was introduced in earlier work~\cite{JS08, JS2, FJS1,
  MSpaper}. The structure of the family allows for efficient search. In
addition, we can establish that the signature family is appropriately ``dense'',
thereby by restricting our search to the signature family, we are not losing
much. We now quickly recall the definition of the signature family:

\medskip
\begin{quote}
  {\bf Signature family.} A distribution (choice model) $\lambda$ is said to
  belong the signature family if for each permutation $\sigma$ that is in the
  support (i.e., $\lambda(\sigma) > 0$) there exist an pair $i$, $j$ such that
  $\sigma(i) = j$ and $\sigma'(i) \neq j$ for any permutation $\sigma'$ in the
  support. Equivalently, for every permutation $\sigma$ in the support of
  $\lambda$, there exists a pair $i$, $j$ such that $\sigma$ ranks $i$ at
  position $j$, but no other permutation in the support ranks $i$ at position
  $r$.
\end{quote} 
\medskip

The above definition states that each element in the support of $\lambda$ has
its `signature' in the data. Before we describe our answers to the questions above, we
introduce two parametric models that we make use of later. 

\subsection{Multinomial Logit (MNL) model} Here we describe the version of the
model as introduced by Luce and Plackett~\cite{plackett, luce}. This is a parametric model with $N$ 
positive valued parameters, one each associated with each of the $N$ items. 
Let $w_i > 0$ be parameter associated with item $i$. Then the probability of
permutation $\sigma \in S_N$ is given by (for example, see~\cite{Marden95})
\begin{align}\label{eq:MNL}
\mathbb{P}_{w}(\sigma) & = \prod_{j=1}^{N}\frac{w_{\sigma^{-1}(j)}}{w_{\sigma^{-1}(j)}+w_{\sigma^{-1}(j+1)}+\dotsb+w_{\sigma^{-1}(N)}}.
\end{align}
Above, $\sigma^{-1}(j) = i$ if $\sigma(i) = j$.

\subsection{Exponential family model}

Now we describe an exponential family of distributions over permutations. 
The exponential family is parametrized by $N^{2}$ parameters $\theta_{ij}$ for 
$1\leq i,j \leq N$. Given such a vector of parameters $\theta$, the probability 
of a permutation $\sigma$ is given by 
\begin{align}\label{eq:EXP}
\mathbb{P}_{\theta}(\sigma) & \propto\exp\left(\sum_{1\leq i,j\leq N}\theta_{ij}\sigma_{ij}\right),\nonumber \\
                                                & = \frac{1}{Z(\theta)} \exp\left(\sum_{1\leq i,j\leq N}\theta_{ij}\sigma_{ij}\right), 
\end{align}
where $Z(\theta) = \sum_{\sigma \in S_N} \exp\left(\sum_{1\leq i,j\leq N}\theta_{ij}\sigma_{ij}\right)$; 
$\sigma_{ij}=1$ iff $\sigma(i)=j$ and $\sigma_{ij}=0$ otherwise. It is well known that with respect to the space
of all first-order marginal distributions, the above described exponential family is dense. Specifically, 
for any doubly stochastic matrix (the first-order marginals) $M = [M_{ij}]$ with $M_{ij} > 0$ for all $i, j$,
there exists $\theta \in \mathbb{R}^{N\times N}$ so that the first-order marginal induced by 
the corresponding exponential family is precisely $M$. An interested reader is referred to, for 
example, monograph~\cite{JW08} for details on this correspondence between parameters of the exponential family and its moments.

\section{Main results} \label{sec:results} 

As our main results, we provide answers to the two questions raised above. We
provide the answers to each of the questions in turn.

\medskip 
\noindent {\em On Question 1(Sparse Approximation):} As our first result, we
establish that given {\em any} doubly stochastic matrix $D$ and $\beps > 0$,
there exists a model $\lambda$ with sparsity $O(N/\beps^2)$ such that
$\normtwo{M(\lambda) - D} \leq \beps$. Thus, we show that by allowing a
``small'' error of $\beps$, one can obtain a significant reduction from
$\Theta(N^2)$ to $O(N/\beps^2)$ in the sparsity of the model that is needed to
explain the observations. More precisely, we have the following theorem.
\begin{theorem}\label{thm:one}
  For any doubly stochastic matrix $D$ and $\beps \in (0,1)$, there exists a
  choice model $\lambda$ such that $\normzero{\lambda} = O(N/\beps^2)$ and
  $\normtwo{M(\lambda) - D} \leq \beps$.
\end{theorem}
We emphasize here that this result holds for {\em any} doubly stochastic matrix
$D$. In such generality, this result is in fact tight in terms of the dependence
on $N$ of the required sparsity. To see that, consider the uniform doubly
stochastic matrix $D$ with all of its entries equal to $1/N$. Then, any choice
model $\lambda$ with $o(N)$ support can have at most $N \times o(N) = o(N^2)$ non-zero
entries, which in turn means that the $\ell_2$ error $\normtwo{M(\lambda) - D}$
is at least $\sqrt{(N^2-o(N^2))/N^2} \approx 1$ for large $N$.

The result of Theorem~\ref{thm:one} also justifies why convex relaxations don't
have any bite in our setting. Specifically, suppose we are given a doubly
stochastic matrix $D$ and a tolerance parameter $\beps > 0$. Then, all the
consistent choice models $\lambda$, which satisfy $\normzero{M(\lambda) - D}
\leq \beps$, have the same $\ell_1$ norm. We claim that ``most'' of such
consistent models $\lambda$ have sparsity $\Theta(N^2)$. More precisely,
following the arguments presented in the proof of~\cite[Theorem 1]{MSpaper}, we
can show that the set of doubly stochastic matrices $\tilde{D}$ such that
$\normtwo{\tilde{D} - D} \leq \beps$ and can be written as $M(\lambda) =
\tilde{D}$ for some model $\lambda$ with sparsity $K < (N-1)^2$ has an $(N-1)^2$
dimensional volume of zero. It thus follows that picking an arbitrary consistent
model $\lambda$ will most certainly yield a model with sparsity $\Theta(N^2)$;
this is a factor $N$ off from the sparsest solution, which has a sparsity of
$O(N)$ (ignoring the $\beps$ dependence).

\medskip
\noindent{\em On Question 2 (Efficient Algorithms):} We now consider the question of efficiently
solving the program in~\eqref{eq:P0}. As explained above, a
brute-force search for a model of sparsity $K$ that is consistent with the data
requires searching over $\exp(\Theta(KN\log N))$ options. We now show that by
restricting ourselves to a reasonably large class of choice models, we can
improve the running time complexity to $O(\exp(\Theta(K \log N)))$ --
effectively eliminating a factor of $N$ from the exponent. More precisely, we
can establish the following result.
\begin{theorem}\label{thm:three}
  Given a noisy observation $D$ and $\beps \in (0, 1/2)$, suppose there exists a
  choice model $\lambda$ in the signature family such that $\normzero{\lambda} =
  K$ and $\normtwo{D - M(\lambda)} \leq \beps$. Then, with a running time
  complexity of $\exp\big(\Theta(K \log N)\big)$, we can find a choice model
  $\lambdahat$ such that $\normzero{\lambdahat} = O(\beps^{-2} K \log N)$ and
  $\normtwo{M(\lambdahat) - D} \leq 2\beps$.
\end{theorem}
Several remarks are in order. The proof of Theorem~\ref{thm:three} is
constructive in the sense that it proposes an algorithm to find a sparse model
with the stated guarantees. The result of Theorem~\ref{thm:three} essentially
establishes that as long as there is a sparse choice model of sparsity $K$ in
the signature family that is an $\beps$-fit to the observations $D$, we can
shave off a factor of $N$ in the exponent from the running time complexity at
the cost of finding a model with sparsity that is essentially within a factor of
$\log N$ of $K$. In other words, we can obtain an exponential reduction in the
running time complexity at the cost of introducing a factor of $\log N$ in the
sparsity.

It is worth pausing here to understand how good (or bad) the computation cost of
$\exp\big(\Theta(K \log N)\big)$ is.  As discussed below (in
Theorem~\ref{thm:two}), for a large class of choice models, the sparsity $K$
scales as $O(\beps^{-2} N)$, which implies that the computation cost scales as
$\exp\big(\Theta(N \log N)\big)$ (ignoring $\beps$ to focus on dependence on
$N$). That is, the computational cost is polynomial in $N! = \exp\big(\Theta(N
\log N)\big)$, or equivalently, polynomial in the dimension of the ambient
space. To put this in perspective, the scaling we obtain is very similar to the
scaling obtained in the recently popular compressive sensing literature, where
sparse models are recovered by solving linear or convex programs, which result
in a computational complexity that is polynomial in the ambient dimension.

Finally, the guarantee of Theorem~\ref{thm:three} is conditional on the
existence of a sparse choice model in the signature family that is an
$\beps$-fit to the data. It is natural to wonder if such a requirement is
restrictive. Specifically, 
% It is natural to wonder if such is indeed the case for
% a general double stochastic matrix $M$. Specifically, 
given any doubly stochastic matrix $D$, there are two possibilities. Firstly, it
may be the case that there is no model in the signature family that is an
$\beps$-fit to the data; in such a case, we may have to lose precision by
increasing $\beps$ in order to find a model in the signature family. Secondly,
even if there did exist such a model, it may not be ``sparse enough''; in other
words, we may end up with a solution in the signature family whose sparsity
scales like $\Theta(N^2)$. Our next result shows that both scenarios described
above do not happen; essentially, it establishes that the signature family of
models is ``dense'' enough so that for a ``large'' class of data vectors, we can
find a ``sparse enough'' model in the signature family that is an $\beps$-fit to
the data.
% In other words, is the signature family ``dense'' enough that for a ``large''
% class of data vectors, we can find a sparse enough choice model in it that is an
% $\beps$-fit to the data? 
More specifically, we can establish that the signature family is ``dense'' as
long as the observations are generated by an MNL model or an exponential family model.

\begin{theorem}\label{thm:two}
  Suppose $D$ is a noisy observation of first-order marginal $M(\lambda)$ with
  $\normtwo{D - M(\lambda)} \leq \beps$ for some $\beps \in (0,1/2)$ and choice
  model $\lambda$ such that
  \begin{itemize}
  \item[1.] either, $\lambda$ is an MNL model with parameters $w_1,\dots, w_N$
    (and without loss of generality $w_1 < w_2 < \ldots < w_N$) such that
    \begin{align}\label{eq:MNL-C}
      \frac{w_N}{\sum_{k=1}^{N-L} w_k} & \leq \frac{\sqrt{\log N}}{N},
    \end{align}
    for $L = N^{\delta}$ for some $\delta \in (0,1)$;
  \item[2.] or, $\lambda$ is an exponential family model with parameters
    $\theta$ such that for any set of four distinct tuples of integers
    $(i_{1},j_{1}),$$(i_{2,},j_{2})$, $(i_{3,}j_{3})$, and $(i_{4},j_{4})$ (with
    $1 \leq i_k, j_k \leq N$ for $1\leq k\leq 4$)
    \begin{align}\label{eq:EXP-C}
      \frac{\exp\left(\theta_{i_{1}j_{1}}+\theta_{i_{2}j_{2}}\right)}{\exp\left(\theta_{i_{3}j_{3}}+\theta_{i_{4}j_{4}}\right)}
      & \leq \sqrt{\log N}.
    \end{align}
  \end{itemize}
  Then, there exists a $\hat{\lambda}$ in the signature family such that:
  $\normtwo{D -\hat{\lambda}} \leq 2\beps$ and
  $\normzero{\hat{\lambda}}=O\Big(N/\beps^2\Big)$.
\end{theorem}

\medskip
\noindent{\bf Remark.} The conditions \eqref{eq:MNL-C} and \eqref{eq:EXP-C} can be
further relaxed by replacing $\sqrt{\log N}$ (in both of them) by $C \log N/\beps^2$ for an 
appropriately chosen (small enough) constant $C > 0$. For the clarity of the exposition,
we have chosen a somewhat weaker condition.

We have established in Theorem~\ref{thm:two} that (under appropriate conditions)
the rich families of MNL and exponential models can be approximated by sparse
models in signature families as far as first-order marginals are concerned. Note
that both families induce distributions that are full support. Thus, if the only
thing we care about are first-order marginals, then we can just use sparse
models in the signature family with sparsity only $O(N)$ (ignoring $\beps$
dependence) rather than distributions that have full support. It is also
interesting to note that in Theorem~\ref{thm:one}, we establish the existence of
a sparse model of $O(N/\beps^2)$ that is an $\beps$-fit to the observations. The
result of Theorem~\ref{thm:two} establishes that by restricting to the signature
family, the sparsity scaling is still $O(N/\beps^2)$ implying that we are not
losing much in terms of sparsity by the restriction to the signature family.

% In the next section, we describe the algorithm we propose to solve the program
% in~\eqref{eq:sparsest_approx} efficiently and also prove
% Theorem~\ref{thm:three}. We present the proofs of Theorems~\ref{thm:one} and
% \ref{thm:two} subsequently. 
In the next section we present the proofs of
Theorems~\ref{thm:one}-\ref{thm:two} before we present the results of our
empirical study.

\section{Proofs}\label{sec:proofs}

\subsection{Proof of Theorem \ref{thm:one}}

We prove this theorem using the probabilistic method. Given the doubly
stochastic matrix $D$, there exists a choice model (by Birkhoff-von Neumann's
result) $\lambda$ such that $M(\lambda) = D$. Suppose we draw $T$ permutations
(samples) independently according to the distribution $\lambda$. Let
$\hat{\lambda}$ denote the empirical distribution based on these $T$ samples. We
show that for $T=N/\beps^2$, on average $\|M(\hat{\lambda}) - D\|_2 \leq \beps$.
Therefore, there must exist a choice model with $T = N/\beps^2$ support size
whose first-order marginals approximate $M$ within an $\ell_2$ error of $\beps$.

To that end, let $\sigma_{1},\sigma_{2},\dotsc,\sigma_{T}$ denote the $T$
samples of permutations and $\hat{\lambda}$ be the empirical distribution (or
choice model) that puts $1/T$ probability mass over each of the sampled
permutations.
Now consider a pair of indices $1\leq i, j \leq N$.  Let $X_{ij}^{t}$ denote the
indicator variable of the event that $\sigma_{t}(i)=j$.  Since the permutations
are drawn independently and in an identically distributed manner, $X_{ij}^{t}$
are independent and identically distributed (i.i.d.) Bernoulli variables for
$1\leq t\leq T$. Further,
\[
\Pb(X^t_{ij} = 1) ~=~\E[X^t_{ij}] ~=~ D_{ij}.
\]
Therefore, the $(i,j)$ component of the first-order marginal $M(\hat{\lambda})$
of $\hat{\lambda}$ is the empirical mean of a Binomial random variable with
parameters $T$ and $D_{ij}$, denoted by $B(T, D_{ij})$. Therefore, with respect
to the randomness of sampling,
\begin{align}
  \E\left[\Big(\frac{1}{T}\sum_{t=1}^T X_{ij}^t - D_{ij}\Big)^2\right] & = \frac{1}{T^2} \Var\Big(B(T,D_{ij})\Big) \nonumber \\
  & = \frac{1}{T^2} T D_{ij} (1-D_{ij}) \nonumber \\
  & \leq \frac{D_{ij}}{T},
\end{align}
where we used the fact that $D_{ij} \in [0,1]$ for all $1\leq i,j \leq N$.
Therefore, 
\begin{align}
\E\Big[ \|M(\hat{\lambda}) - D\|_2^2 \Big]  & = \E\Big[ \sum_{ij} \Big(\frac{1}{T}\sum_{t=1}^T X_{ij}^t - D_{ij}\Big)^2\Big] \nonumber \\
                     & \leq \sum_{ij} \frac{D_{ij}}{T} \nonumber \\ 
                     & = \frac{N}{T}, \label{eq:SP1}
\end{align}
where the last equality follows from the fact that $D$ is a doubly stochastic
matrix and hence its entries sum up to $N$.  From~\eqref{eq:SP1}, it follows
that by selecting $T = N/\beps^2$, the error in approximating the first-order
marginals, $\|M(\hat{\lambda}) - D\|_2$, is within $\beps$ on
average. Therefore, the existence of such a choice model follows by the
probabilistic method. This completes the proof of Theorem~\ref{thm:one}.

\subsection{Proof of Theorem \ref{thm:two}} We prove Theorem \ref{thm:two} using
the probabilistic method as well. As before, suppose that we observe $D$, which
is a noisy version of the first-order marginal $M(\lambda)$ of the underlying
choice model $\lambda$. As per the hypothesis of Theorem~\ref{thm:two}, we shall
assume that $\lambda$ satisfies one of the two conditions: either it is from MNL
model or from exponential family with regularity condition on its parameters.
For such $\lambda$, we establish the existence of a sparse choice model
$\hat{\lambda}$ that satisfies the signature conditions and approximates
$M(\lambda)$ (and hence approximates $D$) well.

As in the proof of Theorem~\ref{thm:one}, consider $T$ permutations
drawn independently and in an identical manner from distribution $\lambda$. 
Let $\hat{\lambda}$ be the empirical distribution of these $T$ samples as
considered before. Following arguments there, we obtain (like~\eqref{eq:SP1}) 
that
\begin{align}
\E\Big[ \|M(\hat{\lambda}) - M(\lambda)\|_2^2 \Big]  & \leq \frac{N}{T}, \label{eq:SP2}
\end{align}
For the choice of $T = 4 N/\beps^2$, using Markov's inequality, we can write
\begin{align}
\Pb\Big(\|M(\hat{\lambda}) - M(\lambda)\|_2^2 \geq \beps^2\Big) & \leq \frac{1}{4}. \label{eq:SP3}
\end{align}
Since $\|M(\lambda) - D\|_2 \leq \beps$, it follows that $\|M(\hat{\lambda}) - D\|_2 \leq 2\beps$ with
probability at least $3/4$. 

Next, we show that the $\hat{\lambda}$ thus generated satisfies the signature
condition with a high probability (at least $1/2$) as well. Therefore, by union
bound we can conclude that $\hat{\lambda}$ satisfies the properties claimed by
Theorem~\ref{thm:two} with probability at least $1/4$.

To that end, let $E_t$ be the event that $\sigma_t$ satisfies the signature
condition with respect to set $(\sigma_1,\dotsc, \sigma_T)$. Since all
$\sigma_1,\dotsc,\sigma_T$ are chosen in an i.i.d. manner, the probability of
all the events are identical. We wish to show that $\Pb\big(\cup_{1\leq t\leq T}
E_t^c\big) \leq 1/2$. This will follow from establishing $T \Pb(E_1^c) \leq
1/2$. To establish this, it is sufficient to show that $\Pb(E_1^c) \leq 1/N^2$
because $T = 4N/\beps^2$.

To that end, suppose $\sigma_1$ is such that $\sigma_1(1) = i_1,\dotsc,
\sigma_1(N) = i_N$. Let $F_j = \{\sigma_t(j) \neq i_j, ~2\leq t\leq T\}$. Then
by the definition of the signature condition, it follows that \[ E_1 = \cup_{j=1}^N F_j.
\] Therefore, \begin{align}
  \Pb\Big(E_{1}^{c}\Big) & = \Pb\Big(\bigcap_{j=1}^{N} F_{j}^{c}\Big) \nonumber \\
                                      & \leq \Pb\Big(\bigcap_{j=1}^{L} F_{j}^{c}\Big) \nonumber \\
                                      & = \Pb\Big(F_{1}^{c}\Big) \prod_{j=2}^{L}\Pb\Big(F^c_{j} \Big| \bigcap_{\ell=1}^{j-1} F^c_{\ell} \Big). \label{eq:UNI}
\end{align}
We will establish that the right hand side of \eqref{eq:UNI} is bounded above by
$O(1/N^2)$ and hence $T\Pb(E_1^c) = O(\beps^{-2}/N) \ll 1/2$ for $N$ large
enough as desired. To establish this bound of $O(1/N^2)$ under two different
conditions stated in Theorem \ref{thm:two}, we consider in turn the two cases:
(i) $\lambda$ belongs to the MNL family with condition~\eqref{eq:MNL-C}
satisfied, and (ii) $\lambda$ belongs to the max-ent exponential family model
with the condition~\eqref{eq:EXP-C} satisfied.
 % we next present the proof for two cases sequentially: (i) under MNL model with
% condition \eqref{eq:MNL-C}, and (ii) under exponential familyt model with condition 
% \eqref{eq:EXP-C}. 

\medskip
\noindent
{\em Bounding \eqref{eq:UNI} under MNL model with~\eqref{eq:MNL-C}.} 
Let $L = N^\delta$ for some $\delta > 0$ as in the hypothesis of Theorem~\ref{thm:two} under
which \eqref{eq:MNL-C} holds. Now 
\begin{align}
\Pb\Big(F_1^c\Big) & = 1 - \Pb\Big(F_1\Big) \nonumber \\
                                & = 1 - \Pb\Big(\sigma_t(1)\neq i_1;~2\leq t\leq T\Big) \nonumber \\
                                 & = 1 - \Pb\Big(\sigma_2(1) \neq i_1\Big)^{T-1} \nonumber \\
                                & = 1- \Big(1-\frac{w_{i_1}}{\sum_{k=1}^N w_k}\Big)^{T-1}. 
\end{align}
For $j \geq 2$, in order to evaluate $\Pb\big(F_j^c | \cap_{\ell=1}^{j-1}F_{\ell}^c\big)$, we
shall evaluate $1 - \Pb\big(F_j | \cap_{\ell=1}^{j-1}F_{\ell}^c\big)$. To evaluate
$\Pb\big(F_j | \cap_{\ell=1}^{j-1}F_{\ell}^c\big)$,  note that the conditioning 
event $\cap_{\ell=1}^{j-1}F_{\ell}^c$ suggests that for each $\sigma_t,~2\leq t\leq T$, 
some assignments (ranks) for the first $j-1$ items are given and we need to find the probability
that $j$th item of each of the $\sigma_2,\dots, \sigma_T$ are not mapped to $i_j$. 
Therefore, given $\cap_{\ell=1}^{j-1}F_{\ell}^c$, the probability that $\sigma_2(j)$ does
map to $i_j$ is $w_j/(\sum_{k \in X} w_k)$, where $X$ is the set of $N-j+1$ elements
that does not include the $j-1$ elements to which $\sigma_2(1),\dots,\sigma_2(j-1)$
are mapped to. Since by assumption (without loss of generality), $w_1 < \dotsb < w_N$, it 
follows that $\sum_{k \in X} w_k \geq \sum_{k=1}^{N-j+1} w_k$. Therefore, 
\begin{align}
\Pb\Big(F_j \Big| \bigcap_{\ell=1}^{j-1} F_\ell^c\Big) & \geq \Big(1-\frac{w_{i_j}}{ \sum_{k=1}^{N-j+1} w_k}\Big)^{T-1}. 
\end{align}
Therefore, it follows that 
\begin{align}
  \Pb\Big(E_{1}^{c}\Big) & \leq \prod_{j=1}^L \Big[1- \Big(1- \frac{w_{i_j}}{\sum_{k=1}^{N-j+1} w_k}\Big)^{T-1}\Big] \nonumber \\
                                      & \leq  \prod_{j=1}^L \Big[1- \Big(1- \frac{w_{N}}{\sum_{k=1}^{N-j+1} w_k}\Big)^{T-1}\Big]\nonumber \\
                                      & \leq \prod_{j=1}^L \Big[1- \Big(1- \frac{w_{N}}{\sum_{k=1}^{N-L+1} w_k}\Big)^{T-1}\Big] \nonumber \\
                                      & = \Big[1- \Big(1- \frac{w_{N}}{\sum_{k=1}^{N-L+1} w_k}\Big)^{T-1}\Big]^L. 
\end{align}                                     
Let $W(L, N) = w_N/(\sum_{k=1}^{N-L+1} w_k )$. 
By hypothesis of Theorem~\ref{thm:two}, it follows that
$W(L,N) \leq \sqrt{\log N}/N$ and $L = N^{\delta}$. Therefore, from above it follows that
\begin{align}
\Pb\Big(E_{1}^{c}\Big) & \leq \Big[1- \Big(1- \frac{\sqrt{\log N}}{N}\Big)^{T-1}\Big]^L \nonumber \\
                                      & \leq \Big[1- \Theta\Big(\exp\Big(-\frac{T\sqrt{\log N}}{N}\Big)\Big]^L, 
\end{align}
where we have used the fact that $1-x = \exp(-x) (1+O(x^2))$ for $x \in [0,1]$ (with $x = \sqrt{\log N}/N$)
and since $T = N/\beps$, $(1+O(\log N/N^2))^T = 1 + o(1) = \Theta(1)$.  Now 
\begin{align}
\exp\Big(-\frac{T\sqrt{\log N}}{N}\Big) & = \exp\Big(-4\sqrt{\log N}/\beps^2\Big) ~\ll ~1. 
\end{align}
Therefore, using the inequality $1-x \leq \exp(-x)$ for $x \in [0,1]$, we have 
\begin{align}
\Pb\Big(E_{1}^{c}\Big) & \leq \exp\Big(-L \exp(-4\sqrt{\log N}/\beps^2)\Big). 
\end{align}
Since $L = N^{\delta}$ for some $\delta > 0$ and $\exp(-4\sqrt{\log N}/\beps^2) = o(N^{\delta/2})$ 
for any $\delta > 0$, it follows that 
\begin{align}
\Pb\Big(E_{1}^{c}\Big) & \leq \exp\Big(-\Theta\big(N^{\delta/2}\big)\Big) \nonumber \\
                                      & \leq O(1/N^2). 
\end{align}
Therefore, it follows that all the $T$ samples satisfy the signature condition with respect to
each other with probability at least $O(1/N) \leq 1/4$ for $N$ large enough. Therefore, we 
have established the existence of desired sparse choice model in signature family. This completes
the proof of Theorem~\ref{thm:two} under MNL model with condition~\eqref{eq:MNL-C}. 

\medskip
\noindent
{\em Bounding \eqref{eq:UNI} under exponential family model with \eqref{eq:EXP-C}.} 
As before, let $L = N^\delta$ for some $\delta > 0$ (the choice of $\delta > 0$ here is arbitrary; for 
simplicity, we shall think of this $\delta$ as being same as that used above).  Now 
\begin{align}
\Pb\Big(F_1^c\Big) & = 1 - \Pb\Big(F_1\Big) \nonumber \\
                                & = 1 - \Pb\Big(\sigma_t(1)\neq i_1;~2\leq t\leq T\Big) \nonumber \\
                                 & = 1 - \Pb\Big(\sigma_2(1) \neq i_1\Big)^{T-1}. \label{eq:exp1}
\end{align}
To bound the right hand side of \eqref{eq:exp1}, we need to carefully understand the implication
of \eqref{eq:EXP-C} on the exponential family distribution.  To start with, suppose parameters
$\theta_{ij}$ are equal for all $1\leq i, j\leq N$. In that case, it is easy to see that all
permutations have equal ($1/N!$) probability assigned and hence the probability 
$\Pb(\sigma_2(1) \neq i_1)$ equals $1-1/N$. However such an evaluation (or bounding) 
is not straightforward as the form of exponential family involves the `partition' function.
To that end, consider $1 \leq i \neq i' \leq N$. Now by definition of exponential family (and
$\sigma_2$ is chosen as per it),
\begin{align}
\Pb\Big(\sigma_2(1) = i \Big) & = \frac{1}{Z(\theta)} \Big[\sum_{\sigma \in S_N(1\to i)} \exp\big(\sum_{kl} \theta_{kl} \sigma_{kl} \big) \Big] \nonumber \\
                                              & = \frac{\exp\big(\theta_{1i} \big)}{Z(\theta)} \Big[\sum_{\sigma \in S_N(1\to i)} \exp\big(\sum_{k \neq 1, l} \theta_{kl} \sigma_{kl} \big) \Big].
\end{align}
In above $S_N(1\to i)$ denotes the set of all permutations in $S_N$ that map $1$ to $i$: 
\[
S_N(1\to i) = \Big\{ \sigma \in S_N: \sigma(1) = i \Big\}.
\]
Given this, it follows that 
\begin{align}
\frac{\Pb\Big(\sigma_2(1) = i \Big)}{\Pb\Big(\sigma_2(1) = i' \Big)} & = \frac{\exp(\theta_{1i})  \Big[\sum_{\sigma \in S_N(1\to i)} \exp\big(\sum_{k \neq 1, l} \theta_{kl} \sigma_{kl} \big) \Big]}{\exp(\theta_{1i'})  \Big[\sum_{\rho \in S_N(1\to i')} \exp\big(\sum_{k \neq 1, l} \theta_{kl} \rho_{kl} \big) \Big]}. \label{eq:exp2}
\end{align}
Next, we will consider a one-to-one and onto map from $S_N(1\to i)$ to $S_N(1\to i')$ (which 
are of the same cardinality). Under this mapping, suppose $\sigma \in S_N(1\to i)$ is 
mapped to $\rho \in S_N(1\to i')$. Then we shall have that 
\begin{align}
\exp\Big(\sum_{kl} \sigma_{kl} \theta_{kl}\Big) & \leq \sqrt{\log N} \exp\Big(\sum_{kl} \rho_{kl} \theta_{kl}\Big). \label{eq:exp3}
\end{align}
This, along with \eqref{eq:exp2} will imply that 
\begin{align}
\frac{\Pb\Big(\sigma_2(1) = i \Big)}{\Pb\Big(\sigma_2(1) = i' \Big)} &\leq \sqrt{\log N}.
\end{align}
This in turn implies that for any $i$, $\Pb(\sigma_2(1) = i) \leq \sqrt{\log N}/N$, which we shall
use in bounding \eqref{eq:exp1}. 

To that end, we consider the following mapping from $S_N(1\to i)$ to $S_N(1\to
i')$. Consider a $\sigma \in S_N(1\to i)$. By definition $\sigma(1) = i$. Let
$q$ be such that $\sigma(q) = i'$. Then map $\sigma$ to $\rho \in S_N(1\to i')$
where $\rho(1) = i'$, $\rho(q) = i$ and $\rho(k) = \sigma(k)$ for $k \neq 1, q$.
Then, \begin{align}
  \frac{\exp\Big(\sum_{kl} \sigma_{kl} \theta_{kl}\Big)}{\exp\Big(\sum_{kl} \rho_{kl} \theta_{kl}\Big)} & = \frac{\exp\big(\theta_{1i} + \theta_{qi'}\big)}{\exp\big(\theta_{1i'} + \theta_{qi}\big)} \nonumber \\
             & \leq \sqrt{\log N},
\end{align}
where the last inequality follows from condition \eqref{eq:EXP-C} in the statement of Theorem~\ref{thm:two}. 
From the above discussion, we conclude that 
\begin{align}
\Pb\Big(F_1^c\Big) & = 1 - \Pb\Big(\sigma_2(1) \neq i_1\Big)^{T-1} \nonumber \\
& \leq 1 - \Big(1 - \frac{\sqrt{\log N}}{N}\Big)^{T-1}.
\end{align} For $j \geq 2$, in order to evaluate $\Pb\big(F_j^c |
\cap_{\ell=1}^{j-1}F_{\ell}^c\big)$, we evaluate $1 - \Pb\big(F_j |
\cap_{\ell=1}^{j-1}F_{\ell}^c\big)$. To evaluate $\Pb\big(F_j |
\cap_{\ell=1}^{j-1}F_{\ell}^c\big)$, note that the conditioning event
$\cap_{\ell=1}^{j-1}F_{\ell}^c$ suggets that for each $\sigma_t,~2\leq t\leq T$,
some assignments (ranks) for first $j-1$ items are given and we need to find the
probability that the $j$th item of each of the $\sigma_2,\dots, \sigma_T$ are
not mapped to $i_j$. Therefore, given $\cap_{\ell=1}^{j-1}F_{\ell}^c$, we wish
to evaluate (an upper bound on) probability of $\sigma_2(j)$ mapping $i_j$ given
that we know assignments of $\sigma_2(1),\dots, \sigma_2(j-1)$. By the form of
the exponential family, conditioning on the assignments $\sigma_2(1),\dotsc,
\sigma_2(j-1)$, effectively we have an exponential family on the space of
permutations of the remaining $N-j+1$ elements. And with respect to that, we wish to
evaluate bound on the marginal probability of $\sigma_2(j)$ mapping to $i_j$. By
an argument identical to the one used above to show that $\Pb(\sigma_2(1) = i)
\leq \sqrt{\log N}/N$, it follows that
\begin{align}
  \Pb\Big(\sigma_2(j) = i_j | \bigcap F_j^c\Big) & \leq \frac{\sqrt{\log N}}{N-j+1} \nonumber \\
                                                                        & \leq \frac{2\sqrt{\log N}}{N}, 
\end{align}
where we have used the fact that $j \leq L  = N^\delta \leq N/2$ (for $N$ large enough). 
Therefore, it follows that 
\begin{align}
\Pb\Big(E_{1}^{c}\Big) & \leq \Big[1- \Big(1- \frac{2 \sqrt{\log N}}{N}\Big)^{T-1}\Big]^L. \label{eq:exp5}
\end{align}
From here on, using arguments identical to those used above (under MNL model), we conclude
that 
\begin{align}
\Pb\Big(E_{1}^{c}\Big) & \leq \exp\Big(-\Theta\big(N^{\delta/2}\big)\Big) \nonumber \\
                                      & \leq O(1/N^2). 
\end{align}
This completes the proof for max-ent exponential family with condition
\eqref{eq:EXP-C} and hence that of Theorem~\ref{thm:two}.

\subsection{Proof of Theorem \ref{thm:three}}\label{ssec:algo}

We are given a doubly-stochastic observation matrix $D$. Suppose there exists a
choice model $\mu$ such that it satisfies signature condition, $\|\mu\|_0 = K$
and $\|M(\mu)-D\|_2 \leq \beps$. Then, the algorithm we describe below finds a
choice model $\lambdahat$ such that $\|\hat{\lambda}\|_0 = O(\beps^{-2} K\log
N)$, $\|M(\hat{\lambda}) - D \|_\infty \leq 2\beps$ in time $\exp(\Theta(K\log
N))$. This algorithm requires effectively searching over space of choice models
from signature family. Before we can describe the algorithm, we introduce a
representation of the models in the signature family, which allows us reduce the
problem into solving a collection of linear programs (LPs).

\medskip 
\noindent{\bf Representation of signature family.}  We start by developing a
representation of choice models from the signature family that is based on their
first order marginal information. 
% Towards that, let us start by developing representation of choice models from 
% signature family with respect to the first order marginal information. 
All the relevant variables are represented by vectors in $N^2$ dimension. For 
example, the data matrix $D = [D_{ij}]$ is represented as an $N^2$ dimensional
vector with components indexed by tuples for the ease of exposition: $D_{ij}$
will be denoted as $D_{(i,j)}$ and the dimensions will be ordered as per lexicographic
ordering of the tuple, i.e. $(i,j) < (i', j')$ iff $i < i'$ or $i = i'$ and $j < j'$. Therefore,
$D$ in column vector form is 
\[ 
D= [D_{(1,1)} ~D_{(1,2)} \dotsc D_{(1,N)} ~D_{(2,1)}\dots D_{(N,N)}]^T.
\]
In a similar manner, we represent a permutation $\sigma \in S_N$ as a $0$-$1$
valued $N^2$ dimensional vector as $\sigma = [\sigma_{(i,j)}]$ with
$\sigma_{(i,j)} = 1$ if $\sigma(i) = j$ and $0$ otherwise.

Now consider a choice model in the signature family with support $K$. Suppose it
has the support $\sigma^1,\dotsc,\sigma^K$ with their respective probabilities
$p_1,\dots, p_K$. Since the model belongs to the signature family, the $K$
permutations have distinct {\em signature} components. Specifically, for each
$k$, let $(i_k,j_k)$ be the signature component of permutation $\sigma^k$ so
that $\sigma^k(i_k) = j_k$ (i.e. $\sigma^k_{(i_k,j_k)}=1$) but
$\sigma^{k'}(i_k)\neq j_k$ (i.e. $\sigma^{k'}_{(i_k,j_k)}=0$) for all $k' \neq
k, ~1\leq k' \leq K$. Now let $M=[M_{(i,j)}]$ be the first-order marginals of
this choice model. Then, it is clear from our notation that
$M_{(i_k, j_k)} = p_k$ for $1\leq k\leq K$ and for any other $(i,j), ~1\leq i, j
\leq N$,
 $M_{(i,j)}$ is a summation of a
subset of the $K$ values $p_1,\dots, p_K$.

The above discussion leads to the following representation of a choice model from the 
signature family. Each choice model is represented by an $N^2 \times N^2$ matrix
with $0$-$1$ entries, say $Z=[Z_{(i,j) (i',j')}]$ for $1\leq i, j, i',j' \leq N$: in $Z_{(i,j)(i',j')}$, 
$(i,j)$ represents a row index while $(i',j')$ represents a column index. The choice model 
with support $K$ is identified with its $K$ signature components $(i_k,j_k)$, $1\leq k\leq K$. 
The corresponding $Z$ has all $N^2 - K$ columns corresponding to indices other than these $K$ 
tuples equal to $0$. The columns corresponding to the $(i_k, j_k)$ indices, $1\leq k\leq K$, are non-zero with each representing a permutation consistent with the signature
condition: for each $(i_k, j_k)$, $1\leq k\leq K$, 
\begin{align} 
Z_{(i,j) (i_k, j_k)} & \in \{0,1\}, \quad \mbox{for~all}~ 1\leq i, j \leq N, \label{eq:sign1} \\
Z_{(i_k, j_k) (i_k, j_k)} & = 1, \label{eq:sign2}\\ 
Z_{(i,j)(i_k,j_k)} & = 0, \quad \mbox{if}~ (i,j) \in \{(i_{k'},j_{k'}): 1\leq k' \leq K, ~k' \neq k\},\label{eq:sign3} \\ 
\sum_{\ell=1}^N Z_{(i,\ell)(i_k,j_k)} ~= 1, & \quad  \sum_{\ell=1}^N Z_{(\ell,j)(i_k,j_k)} ~= 1, 
~~\mbox{for~all}~1\leq i, j\leq N. \label{eq:sign4}
\end{align}
Observe that \eqref{eq:sign2}-\eqref{eq:sign3} enforce the signature condition
while \eqref{eq:sign4} enforces the permutation structure. In summary, given a
set of $K$ distinct pairs of indices, $(i_k, j_k)$, $1\leq k\leq K$ with $1\leq
i_k, j_k \leq N$, \eqref{eq:sign1}-\eqref{eq:sign4} represent the set of all
possible signature family with these indices as their signature components.

Notice now that given the above representation, the problem of finding a choice model of support $K$ within the signature family that is within an $\beps$-ball of the observed first-order marginal data, $D$, may be summarized as finding a $Z$ satisfying \eqref{eq:sign1}-\eqref{eq:sign4} and in addition, satisfying $\| D - ZD \|_2 \leq \beps$. The remainder of this section will be devoted to solving this problem tractably.

\medskip
\noindent{\bf Efficient representation of signature family.} A signature
family choice model with support $K$ can, in principle, have any $K$ of the
$N^2$ possible tuples as its signature components. Therefore, one way to search
the signature family for choice models is to first pick a set of $K$ tuples (there are ${N^2 \choose K}$ such sets)
and then for that particular set of $K$ tuples, search among
all $Z$s satisfying \eqref{eq:sign1}-\eqref{eq:sign4}. It will be the complexity of this procedure that essentially drives the complexity of our approach. 
%As we shall see, the
%basic searching complexity for determining a sparse choice model arises from
%going over each of the distinct ${N^2 \choose K}$ tuples. 
To this end we begin with the following observation: the problem of optimizing a linear functional of $Z$ subject to the constraints \eqref{eq:sign1}-\eqref{eq:sign4} is equivalent to optimizing the functional over the constraints 
\begin{align} 
Z_{(i,j) (i_k, j_k)} & \in [0,1], \quad \mbox{for~all}~ 1\leq i, j \leq N, \label{eq:sign1a} \\
Z_{(i_k, j_k) (i_k, j_k)} & = 1, \label{eq:sign2a}\\ 
Z_{(i,j)(i_k,j_k)} & = 0, \quad \mbox{if}~ (i,j) \in \{(i_{k'},j_{k'}): 1\leq k' \leq K, ~k' \neq k\},\label{eq:sign3a} \\ 
\sum_{\ell=1}^N Z_{(i,\ell)(i_k,j_k)} ~= 1, & \quad  \sum_{\ell=1}^N Z_{(\ell,j)(i_k,j_k)} ~= 1, 
~~\mbox{for~all}~1\leq i, j\leq N. \label{eq:sign4a}
\end{align}
%This is because the
%the points described by \eqref{eq:sign1}-\eqref{eq:sign4} form the extreme
%points of the following relaxation: for each $(i_k, j_k)$, $1\leq k\leq K$,
%\begin{align} 
%Z_{(i,j) (i_k, j_k)} & \in [0,1], \quad \mbox{for~all}~ 1\leq i, j \leq N, \label{eq:sign1a} \\
%Z_{(i_k, j_k) (i_k, j_k)} & = 1, \label{eq:sign2a}\\ 
%Z_{(i,j)(i_k,j_k)} & = 0, \quad \mbox{if}~ (i,j) \in \{(i_{k'},j_{k'}): 1\leq k' \leq K, ~k' \neq k\},\label{eq:sign3a} \\ 
%\sum_{\ell=1}^N Z_{(i,\ell)(i_k,j_k)} ~= 1, & \quad  \sum_{\ell=1}^N Z_{(\ell,j)(i_k,j_k)} ~= 1, 
%~~\mbox{for~all}~1\leq i, j\leq N. \label{eq:sign4a}
%\end{align}
It is easy to see that the points described by the set of equations
\eqref{eq:sign1}-\eqref{eq:sign4} are contained in the polytope above described
by equations \eqref{eq:sign1a}-\eqref{eq:sign4a}. Thus, in order to justify our observation, it suffices to show that the polytope above is the convex hull of points satisfying \eqref{eq:sign1}-\eqref{eq:sign4}. But this again follows from the Birkhoff-Von Neumann theorem.

\medskip
\noindent{\bf Searching the signature family.} We now describe the main algorithm
that will establish the result of Theorem~\ref{thm:three}. The algorithm
succeeds in finding a choice model $\hat{\lambda}$ with sparsity
$\|\hat{\lambda}\|_0 = O(\beps^{-2} K \log N)$ and error $\|M(\hat{\lambda}) -
D\|_\infty \leq 2\beps$ if there exists a choice model $\mu$ in signature family
with sparsity $K$ that is near consistent with $D$ in the sense that $\|M(\mu) -
D\|_\infty \leq \beps$ (note that $\|\cdot\|_2 \leq \|\cdot\|_\infty$). The
computation cost scales as $\exp\big(\Theta(K \log N)\big)$. Our algorithm uses
the so called {\em Multiplicative Weights} algorithm utilized within the framework developed by Plotkin, Shmoys and Tardos \cite{PST} for fractional
packing (also see \cite{AK}).

The algorithm starts by going over all possible ${N^2 \choose K}$ subsets of
possible signature components in any order till desired choice model
$\hat{\lambda}$ is found or all are exhausted. In the latter case, we declare the infeasibility of finding a $K$ sparse choice model in the signature family that is near
consistent. Now consider any such set of $K$ signature components, $(i_k, j_k)$
with $1\leq k\leq K$. By the definition of the signature family, the values
$D_{(i_k,j_k)}$ for $1\leq k\leq K$ are probabilities of the $K$ permutations in
the support. Therefore, we check if $1-\beps \leq \sum_{k=1}^K D_{(i_k, j_k)}
\leq 1 + \beps$. If not, we reject this set of $K$ tuples as signature components
and move to the next set. If yes, we continue towards finding a choice model
with these $K$ as signature components and the corresponding probabilities.

%The first step is to verify if choice model with these $K$ signature components
%is feasible. As discussed above, we can verify this in $O(KN^{2.5})$ time. 
%[TODO: WE CAN ELIMINATE THIS FIRST STEP]
%If
%such a model indeed exists, we search for the appropriate model. 

The choice
model of interest to us, and represented by a $Z$ satisfying
\eqref{eq:sign1}-\eqref{eq:sign4}, should be such that $D \approx Z D$.
%where
%$Y$ is viewed as an $N^2$ dimensional vector and $Z$ as $N^2 \times N^2$ matrix.
Put another way, we are interested in finding a $Z$ such that
% The choice model of our interest, represented by $Z$ satisfying
% \eqref{eq:sign1}-\eqref{eq:sign4}, should be such that $D \approx Z D$,
% where $D$ is viewed as $N^2$ dimensional vector and $Z$  as 
% $N^2 \times N^2$ matrix. Putting it other way, we are interested in finding
% $Z$ so that 
\begin{align}
D_{(i,j)} - \beps ~&~ \leq \sum_{k=1}^K Z_{(i,j)(i_k,j_k)} D_{(i_k, j_k)} ~\leq~D_{(i,j)} + \beps, ~~\mbox{for~all}~~1\leq i, j \leq N^2 \label{eq:sign5} \\
& \quad~Z ~ \mbox{satisfies ~} \eqref{eq:sign1}-\eqref{eq:sign4}. \label{eq:sign6}
\end{align}
This is precisely the setting considered by Plotkin-Shmoys-Tardos \cite{PST}:
$Z$ is required to satisfy a certain collection of `difficult' linear
inequalities \eqref{eq:sign5} and a certain other collection of `easy' convex
constraints \eqref{eq:sign6} (easy, since these constraints can be replaced by \eqref{eq:sign1a}-\eqref{eq:sign4a} which provide a relaxation with no integrality gap as discussed earlier). If there is a
feasible solution satisfying \eqref{eq:sign5}-\eqref{eq:sign6}, then \cite{PST}
finds a $Z$ that satisfies \eqref{eq:sign5} approximately and \eqref{eq:sign6} exactly. Otherwise, the procedure provides a certificate of the infeasibility of the above program; i.e. a certificate showing that no signature choice model approximately consistent with the data and with the $K$ signature components in question exists. We describe the precise algorithm next.

\newcommand{\cP}{{\mathcal P}}

% For ease of notation, we will denote the choice model matrix $Z$ of dimension
% $N^2 \times N^2$ (effectively $N^2 \times K$) by a vector $z$ of $ K N^2$ dimension; 
% we shall think of \eqref{eq:sign5} as $2N^2$ inequalities denoted as $Az \geq b$ 
% with $A$ being $2N^2 \times KN^2$ matrix and $b$ being $2N^2$ dimensional vector; 
% and \eqref{eq:sign6} by the linear relaxation \eqref{eq:sign1a}-\eqref{eq:sign4a},
% denoted by $\cP$. Thus, we are interested in finding $z \in \cP$ so that $Az \geq b$.  

% The~\cite{PST} framework essentially tries to solve the {\em Lagrangian} relaxation
% of $Az \geq b$ over $z \in \cP$ in an iterative manner. To that end, let $p_\ell$ be
% the Lagrangian variable (or weight) parameter associated with the $\ell$th constraint
% $a_\ell^T z \geq b_\ell$ for $1\leq \ell \leq 2N^2$ (where $a_\ell$ is the $\ell$th row
% of $A$). We shall update them iteratively: let $t \in \{0,1, \dots\}$ represent
% the index of iteration. Initially, $t = 0$ and $p_\ell(0) = 1$ for all $\ell$. Given
% $p(t) = [p_\ell(t)]$, find $z^t$ by solving linear program 

For ease of notation, we denote the choice model matrix $Z$ of dimension $N^2
\times N^2$ (effectively $N^2 \times K$) by a vector $z$ of $ K N^2$ dimension;
we think of \eqref{eq:sign5} as $2N^2$ inequalities denoted by $Az \geq b$ with
$A$ being a $2N^2 \times KN^2$ matrix and $b$ being a $2N^2$ dimensional vector. Finally, the set of $z$ satisfying \eqref{eq:sign6}, is denoted $\cP$. Thus, we are
interested in finding $z \in \cP$ such that $Az \geq b$.

The framework in \cite{PST} essentially tries to solve the {\em Lagrangian}
relaxation of $Az \geq b$ over $z \in \cP$ in an iterative manner. To that end,
let $p_\ell$ be the Lagrangian variable (or weight) parameter associated with
the $\ell$th constraint $a_\ell^T z \geq b_\ell$ for $1\leq \ell \leq 2N^2$
(where $a_\ell$ is the $\ell$th row of $A$). We update the weights iteratively:
let $t \in \{0,1, \dotsc\}$ represent the index of the iteration. Initially, $t = 0$
and $p_\ell(0) = 1$ for all $\ell$. Given $p(t) = [p_\ell(t)]$, we find $z^t$ by
solving the linear program
\begin{align}\label{eq:mw1}
{\sf maximize} & ~\sum_{\ell} p_\ell(t) (a_\ell^T z - b_\ell) \nonumber \\ 
{\sf over} & ~~z \in {\rm co}(\cP).  
\end{align}
Notice that by our earlier discussion, ${\rm co}(\cP)$ is the polyhedron defined by the linear inequalities \eqref{eq:sign1a}-\eqref{eq:sign4a}, so that optimal {\em basic} solutions to the LP above are optimal solutions to the optimization problem obtained if one replaced ${\rm co}(\cP)$ with simply $\cP$. Now in the event that the above LP is infeasible, or else if its optimal value is negative, we declare immediately that there does not exist a $K$-sparse choice model with the $K$ signature components in question that is approximately consistent with the observed data; this is because the above program is a relaxation to $\eqref{eq:sign5}-\eqref{eq:sign6}$ in that $\eqref{eq:sign5}$ has been relaxed via the `lagrange' multiplier $p(t)$. Further, if the original program were feasible, then our LP should have a solution of non-negative value since the weights $p(t)$ are non-negative. Assuming, we do {\em not} declare infeasibility, the solution $z_t$ obtained is a $K$-sparse choice model whose signature components correspond to the $K$ components we began the procedure with. 

%In the event that the program is infeasible, we can declare immediately that there does not exist a $K$-sparse choice model with the $K$ signature components in question that is approximately consistent with the observed data; this is because the above program is a relaxation to $\eqref{eq:sign5}-\eqref{eq:sign6}$ in that $\eqref{eq:sign5}$ has been relaxed via the `lagrange' multiplier $p(t)$.  

%We insist on $z^t$ being an extreme point of $\cP$. Thus, even in case there are
%multiple solutions, $z^t$ will be integral (its components will be $0$ or $1$)
%corresponding to a $K$-sparse choice model in the signature family. 

Assuming that the linear program is feasible, and given an optimal basic feasible solution $z_t$, the weights $p(t+1)$ are obtained as follows: for $\delta =
\min\big(\beps/8, 1/2\big)$, we set: 
\begin{align}\label{eq:mw2}
p_{\ell}(t+1) & = p_{\ell} \big(1 - \delta(a_\ell^T z^t - b_\ell)\big). 
\end{align}
The above update \eqref{eq:mw2} suggests that if the $\ell$th inequality is not
satisfied, we should increase the penalty imposed by $p_\ell(t)$ (in proportion to the degree of violation) or else, if it is satisfied, we decrease the penalty imposed by $p_l(t)$ in proportion to the `slack' in the constraint. Now, $a_\ell^T
z^t - b_\ell \in [-2,2]$. To see this note that: First, $b_\ell \in [0,1]$ since it corresponds to an entry in a
non-negative doubly stochastic matrix $D$. Further, $a_\ell^T z^t \in [0, 1+\beps]$
since it corresponds to the summation of a subset of $K$ non-negative entries
$D_{(i_k,j_k)}, ~1\leq k\leq K$ and by choice we have made sure that the
sum of these $K$ entries is at most $1+\beps$. Hence, the multiplicative update to each of the
$p_\ell(\cdot)$ is by a factor of at most $(1\pm 2\delta)$ in a single iteration. Such a bound on the relative change of these weights is necessary for the success of the algorithm. 

Now, assume we have not declared infeasibility for all $t \leq T$ and consider the sequence of solutions, $z^t$. Further, set $T = 64 \beps^{-2} \ln (2N^2) = O(\beps^{-2} \log N)$, and define $\hat z = \frac{1}{T} \sum_{t=0}^{T-1} z^t$. Then, we have via Corollary 4 in \cite{AK} (see also, \cite[Section 3.2]{AK}), that
\begin{align}
a_\ell^T \hat{z} & \geq b_\ell - \beps, ~~\mbox{for~all} ~1\leq \ell \leq 2N^2. \label{eq:mw3}
\end{align}

%Now consider the sequence of $z^t$ produced for $t \leq T$ where $T = O(\beps^{-2} \log N)$
%(precisely, $T = 64 \beps^{-2} \ln (2N^2)$ as per \cite[Corollary 4]{AK} and its utilization 
%in \cite[Section 3.2]{AK}). 
%If for any $t$, the value of the objective % CHECK
%of optimization \eqref{eq:mw1} is less than $0$, then we declare infeasibility. 
%This is because, if indeed there was $z$ that was a feasible solution to 
%\eqref{eq:sign5}-\eqref{eq:sign6}, then this optimization problem must 
%have cost of optimal solution $\geq 0$. On the other hand, if indeed for all
%iterations $t\leq T$, this condition is satisfied then 
%$\hat{z} = \frac{1}{T} (\sum_{t=0}^{T-1} z^t)$ is such that 
%(see \cite[Section 3.2]{AK}) % CHECK
%\begin{align}
%a_\ell^T \hat{z} & \geq b_\ell - \beps, ~~\mbox{for~all} ~1\leq \ell \leq 2N^2. \label{eq:mw3}
%\end{align}

Now $\hat{z}$ corresponds to a choice model (call it $\hat \lambda$) with support over at most $O(KT)$ $=$
$O\big(\frac{K}{\beps^2} \log N\big)$ permutations since each $z^t$ is a choice model with
support over $K$ permutations in the signature family. Further, \eqref{eq:mw3} implies that $\|M(\hat{\lambda}) -
D\|_\infty \leq 2\beps$. 

Finally, note that the computational complexity of the above described algorithm, for a given subset of
$K$ signature components is polynomial in $N$. Therefore, the overall
computational cost of the above described algorithm is dominated by term ${N^2
  \choose K}$ which is at most $N^{2K}$. That is, for any $K \geq 1$, the
overall computation cost of the algorithm is bounded above by $\exp\big(\Theta(K
\log N)\big)$. This completes the proof of Theorem \ref{thm:three}.

\medskip
\noindent{\bf Utilizing the algorithm.} It is not clear a priori if for given set of first-order marginal information, $D$, there exists a signature family of sparsity $K$ within 
some small error $\beps > 0$ with $\beps \leq \beps_0$ where $\beps_0$
is the maximum error we can tolerate. The natural way to adapt the above
algorithm is as follows. Search over increasing values of $K$ and for each
$K$ search for $\beps = \beps_0$. For the first $K$ for which the algorithm
succeeds, it may be worth optimizing over the error allowed, $\beps$, by means of a
binary search: $\beps_0/2, \beps_0/4, \dotsc$. Clearly such a procedure
would require $O(\log 1/\beps)$ additional run of the same algorithm for the
given $K$, where $\beps$ is the best precision we can obtain.

\section{An empirical study} \label{sec:empirical}

This Section is devoted to answering the following, inherently empirical, question: 
\newline
\newline
\noindent {\em Can sparse choice models fit to limited information about the underlying `true' choice model be used to effectively uncover information one would otherwise uncover with ostensibly richer data?} 
\newline
\newline
In this section, we describe an empirical study we conducted that supports an affirmative answer to the above question. For the purpose of the study, we used the well-known APA (American
Psychological Association) dataset that was first used by~\cite{diaconis89} in
order to demonstrate the underlying structure one can unearth by studying the
appropriate lower-dimensional `projections' of choice models, which include
first and second order marginals. 

Specifically, the dataset comprises the ballots collected for electing the
president of the APA. Each member expresses her/his preferences by rank ordering the
candidates contesting the election. In the year under consideration, there were
five candidates contesting the election and a total of 5,738 votes that were
complete rankings.  This information yields a distribution mapping each
permutation to the fraction of voters who vote for it.  Given all the votes, the
winning candidate is determined using the {\em Hare} system
(see~\cite{fishburn83} for details about the Hare system).

A common issue in such election systems is that it is a difficult cognitive task
for voters to rank order all the candidates even if the number of candidates is
only five. This, for example, is evidenced by the fact that out of more than 15,000
ballots cast in the APA election, only 5,738 of them are complete. The problem
only worsens as the number of candidates to rank increases. One way to overcome
this issue is to design an election system that collects only partial
information from members. The partial information still retains some of the
structure of the underlying distribution, and the loss of information is the
price one pays for the simplicity of the election process. For example, one can
gather first-order partial information i.e., the fraction of people who rank
candidate $i$ to position $r$. As discussed by~\cite{diaconis89}, the
first-order marginals retain useful underlying structure like: (1) candidate 3
has a lot of ``love'' (28\% of the first-position vote) and ``hate'' (23\% of
the last-position vote) vote; (2) candidate 1 is strong in second position (26\%
of the vote) and low hate vote (15\% of last-position vote); (3) voters seem
indifferent about candidate 5.

Having collected only first order information, our goal will be to answer natural questions such as: who should win the election? or what is the `socially preferred' ranking of candidates? Of course, there isn't a definitive manner in which the above questions might be answered. However, having a complete
distribution over permutations affords us the flexibility of using any of the several rank
aggregation systems available. In order to retain this flexibility, we will fit a sparse distribution to the partial information and then use this sparse distribution as input to the rank aggregation system of choice to determine the `winning' ranking. Such an approach would be of value if the sparse
distribution can capture the underlying structural information of the problem at
hand. Therefore, with an aim to understanding the type of structure sparse models
can capture, we first considered the first-order marginal information of the
dataset (or distribution). We let $\lambda$ denote the underlying ``true''
distribution corresponding to the 5,738 complete rankings of the 5 candidates.
The $5 \times 5$ first-order marginal matrix $D$ is given in
Table~\ref{tab:apa_firstorder}.

\begin{table}
  \centering
  \begin{tabular}{>{\centering\arraybackslash}m{2cm}
      >{\centering\arraybackslash}m{1cm} >{\centering\arraybackslash}m{1cm} 
      >{\centering\arraybackslash}m{1cm} >{\centering\arraybackslash}m{1cm}
      >{\centering\arraybackslash}m{1cm}}
    & \multicolumn{5}{c}{{\bf Rank}} \\
    {\bf Candidate} & {\bf 1} & {\bf 2} & {\bf 3} & {\bf 4} & {\bf 5}\\
    \hline
    1& 18 & 26 & 23 & 17 & 15 \\
    2 & 14 & 19 & 25 & 24 & 18 \\
    3 & 28 & 17 & 14 & 18 & 23 \\
    4 & 20 & 17 & 19 & 20 & 23 \\
    5 & 20 & 21 & 20 & 19 & 20 \\
    \hline
  \end{tabular}
  \caption{The first-order marginal matrix where the entry corresponding to
    candidate $i$ and rank $j$ is the percentage of voters who rank candidate $i$ to position $j$}
  \label{tab:apa_firstorder}
\end{table}

For this $D$, we ran a {\em heuristic} version of the algorithm described in
Section~\ref{ssec:algo}. Roughly speaking, the heuristic tries to find in a {\em
  greedy} manner a sparse choice model in the signature family that approximates
the observed data. It runs very fast (polynomial in $N$) and seems to provide approximations of a quality guaranteed by the algorithm in Section~\ref{ssec:algo}.
However, we are unable to prove any guarantees for it. To keep the exposition
simple, and to avoid distraction, we do not describe the heuristic here but simply refer the interested reader to \cite{JThesis11}. Using the
heuristic, we obtained the following sparse model $\lambdahat$: 
\begin{align*}
  24153 & \quad 0.211990 \\
  32541 & \quad 0.202406 \\
  15432 & \quad 0.197331 \\
  43215 & \quad 0.180417 \\
  51324 & \quad 0.145649 \\
  23154 & \quad 0.062206
\end{align*}
In the description of the model $\lambdahat$ above, we have adopted the notation used
in~\cite{diaconis89} to represent each rank-list by a five-digit number in which
each candidate is shown in the position it is ranked i.e., 24153 represents the
rank-list in which candidate 2 is ranked at position 1, candidate 4 is ranked at
position 2, candidate 1 is ranked at position 3, candidate 5 is ranked at
position 4, and candidate 3 is ranked at position 5. Note that the support size
of $\lambdahat$ is only $6$, which is a significant reduction from the full
support size of $5!  = 120$ of the underlying distribution. The average relative
error in the approximation of $M$ by the first-order marginals
$M(\hat{\lambda})$ is less than $0.075$, where the average relative error is
defined as
\[
\sum_{1\leq i,j \leq 5} \frac{|M(\hat{\lambda})_{ij} - D_{ij}|}{D_{ij}}.
\]
Note that this measure of error, being relative, is more stringent than measuring additive error. The main conclusion we can draw from the small
relative error we obtained is that the heuristic we used can successfully find sparse models that are a good fit to the data in interesting practical
cases.

\subsection{Structural Conclusions}
\begin{figure}
  \label{fig:CDF}
  \centering
  \scalebox{0.55}{\includegraphics{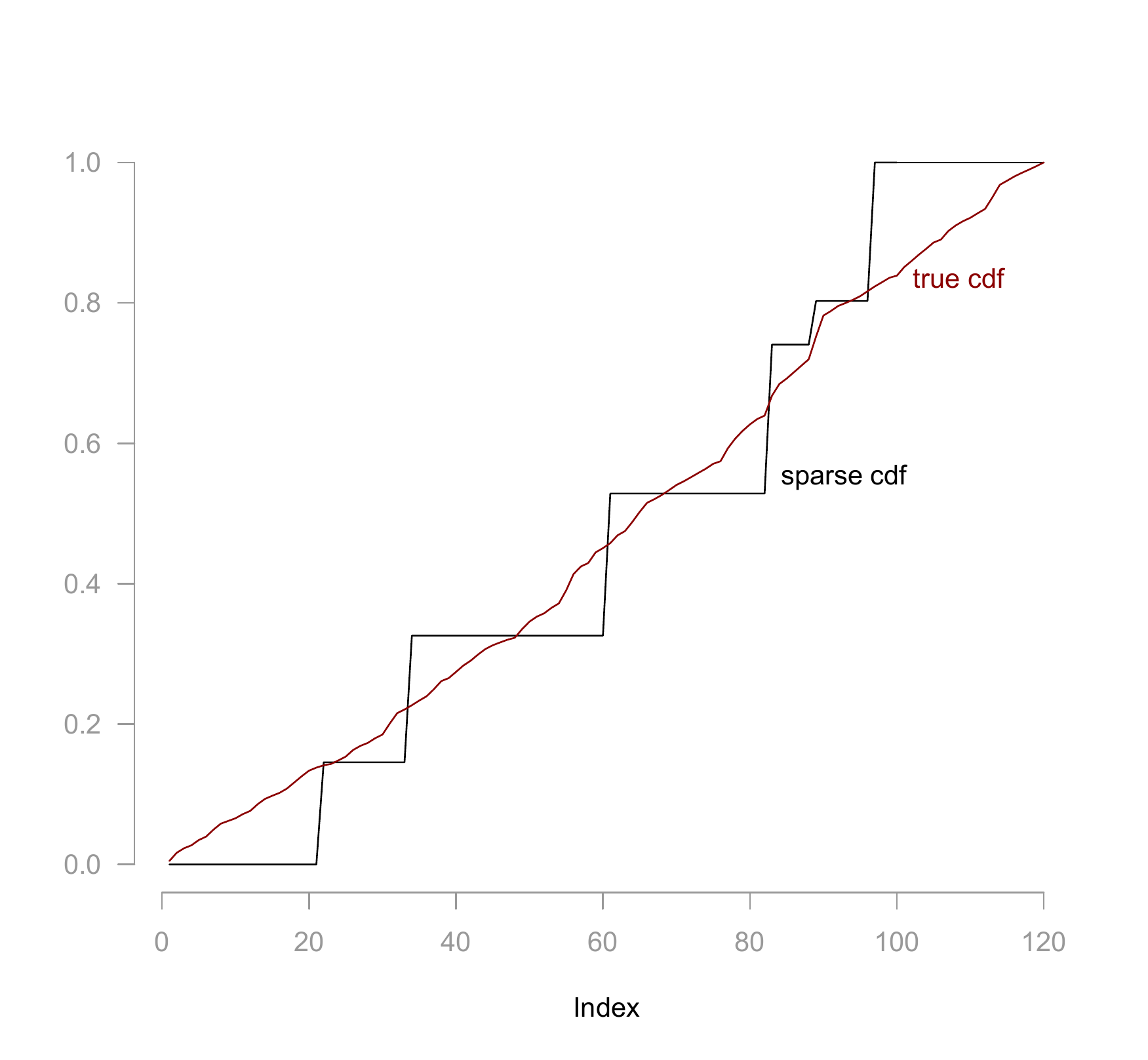}}
  \caption{Comparison of the CDFs of the true distribution and the sparse
    approximation we obtain for the APA dataset. The $x$-axis represents the $5!
    = 120$ different permutations ordered so that nearby permutations are close
    to each other with respect to the pairwise transposition distance.}
\end{figure}

Now that we have managed to obtain a huge reduction in sparsity at the cost of
an average relative error of $0.075$ in approximating first-order marginals, we
next try to understand the type of structure the sparse model is able to capture
from just the first-order marginals. More importantly, we will attempt to compare these conclusions with conclusions drawn from what is ostensibly `richer' data:
\newline
\newline
\noindent {\bf Comparing CDFs: }We begin with comparing the `stair-case' curves of the cumulative
distribution functions (CDF) of the actual distribution $\lambda$ and the sparse
approximation $\hat{\lambda}$ in Figure~\ref{fig:CDF}. Along the $x$-axis in the
plot, the permutations are ordered such that nearby permutations are ``close''
to each other in the sense that only a few transpositions (pairwise swaps) are
needed to go from one permutation to another. The figure visually represents
how well the sparse model approximates the true CDF.

Now, one is frequently interested in a functional of the underlying choice model such as determining a winner, or perhaps, determining a socially preferred ranking. We next compare conclusion drawn from applying certain functionals to the sparse choice model we have learned with conclusions drawn from applying the same functional to what is ostensibly richer data: 
\newline
\newline
\noindent {\bf Winner Determination: }
Consider a functional of the distribution over rankings meant to capture the most `socially preferred' ranking. There are many such functionals, and the {\em Hare system} provides one such example. When applied to the sparse choice model we have learned this yields the permutation 13245. Now, one may use the Hare system to determine a winner with {\em all} of the voting data. This data is {\em substantially} richer than the first order marginal information used by our approach. In particular, it consists of 5,738 votes that consists of entire permutations of the candidates (from which our first order marginal information was derived), and approximately 10,000 additional votes for partial rankings of the same candidates. Applying the Hare system here also yields $1$ as the winning candidate as reported by \cite{diaconis89}.
\newline
\newline
\noindent {\bf Rank Aggregation: }
In addition to determining a winner, the Hare system applied to a choice model also yields an `aggregate' permutation which, one may argue, represent the aggregate opinions of the population in a `fair' way. Now, as reported above, the Hare system applied to our sparse choice model yields the permutation $13245$. As it turns out, this permutation is in remarkable agreement with conclusions drawn by Diaconis using {\em higher} order partial information derived from the same set of $5,738$ votes used here. In particular, using second-order marginal data, i.e. information on the fraction of voters that ranked candidates $\{i,j\}$ to positions $\{k,l\}$ (without accounting for order in the latter set) for all distinct $i,j,k,l$ yields the following conclusion, paraphrased from \cite{diaconis89}: There is a strong effect for candidates $\{1,3\}$ to be ranked first and second and for candidates $\{4,5\}$ to be ranked fourth and fifth, with candidate $2$ in the middle. Diaconis goes on to provide some color to this conclusion by explaining that voting is typically along partisan lines (academicians vs. clinicians) and as such these groups tend to fall behind the candidate groups $\{1,3\}$ and $\{4,5\}$. Simultaneously, these candidate groups also receive `hate' vote wherein they are voted as the least preferred by the voters in the opposing camp. $2$ is apparently something of a compromise candidate. Remarkably, we have arrived at the very same permutation using first order data.  
\newline
\newline
\noindent{\bf Sparse Support Size: }It is somewhat tantalizing to relate the support size ($6$) of the sparse choice model learned with the structure observed in the dataset by Diaconis~\cite{diaconis89} discussed in our last point:
there are effectively three types (groups) of candidates, viz. $\set{2}$, $\set{1, 3}$
and $\set{4,5}$, in the eyes of the partisan voters. Therefore, all votes effectively
exhibit an ordering/preference over these three groups primarily and therefore
effectively the votes are representing $3! = 6$ distinct preferences. This is precisely the size of the support of our sparse approximation; of course, this explanation is not perfect since the permutations in the choice model learned split up these groups.  

\section{Discussion}\label{sec:discussion}

\subsection{Summary}

Choice models are an integral part of various important decision making tasks.
Sparse approximations to the underlying true choice model based on
(information-) limited observed data are particularly attractive as they are
simple and hence easy to integrate into complex decision making tasks. In
addition, learning sparse models from marginal data provides a nonparametric
approach to choice modeling. This paper, in a sense, has taken important steps
towards establishing sparse choice model approximation as a viable option.

As the first main result, we showed that for first-order information, if we
are willing to allow for an $\ell_2$ error of $\beps$ in approximating a given
doubly stochastic matrix (that is, noisy observations of first-order marginal information), 
then there exists a choice model with sparsity
$O(N/\beps)$ that is an $\beps$-fit to the data. Note that this is a significant
reduction from $\Theta(N^2)$ that is guaranteed by the Birhkoff-von Neumann and
Caratheodory's theorems. Given that we can expect to find sparse models, we
considered the issue of efficient recovery of sparse models. We showed that as
long as there is a choice model $\lambda$ of sparsity $K$ in the signature
family that is an $\beps$-fit to the data, we can find a choice model of
sparsity $O(K \log N)$ that is a $2\beps$-fit to the data in time
$O(\exp(\Theta(K \log N)))$ as opposed to the brute-force time complexity of
$O(\exp(\Theta(K N \log N)))$. The computational efficiency is achieved by means
of the `signature condition'. In prior work, this condition was shown to be
useful in learning an already sparse choice model from its noise-free first order marginals. This work
establishes, in a sense, the robustness of these conditions. Finally, we demonstrated the ubiquity of the signature family by showing that it 
%we justified
%the existence of a model in the signature family as a good fit to data by
%showing that the signature family 
is appropriately ``dense'' for a large class
of models.

% As the first main result, we established the universality of the sparse
% approximation: all choice models can be well approximated by sparse choice
% models (in terms of observed first-order marginal data). Next, we showed that
% such sparse model approximation can be learnt computationally efficiently, at
% least for a large sub-class of choice models. The computational efficiency is
% achieved by means of `signature condition'. In the prior work, this condition
% was shown to be useful in learning exact choice model from noise-free data. This
% work establishes, in a sense, robustness of these conditions. More generally, it
% suggests that for a large sub-class of choice models that are well approximated
% by choice models from signature family are easier to learn.

In the recently popular compressive sensing literature, the restricted null
space condition has been shown to be necessary and sufficient for efficient learning
of sparse models via linear programs. It was shown in the past that this restricted null
space condition (or effectively a linear programming relaxation of our problem) is ineffective in learning sparse choice
models. In that sense, this work shows that `signature conditions' are another
set of sufficient conditions that help learn sparse choice models in a computationally
efficient fashion.

\subsection{Beyond first-order marginals}

Here we discuss the applicability of the results of this work beyond first-order marginal information. The proof for the result
(Theorem~\ref{thm:one}) on ``how sparse the sparse models are'' does not exploit
the structure of first-order marginals and hence can be extended in a reasonably
straightforward manner to other types of marginal information. Similarly, we
strongly believe the result (Theorem~\ref{thm:two}) that we can find good
approximations in the signature family for a large class of choice models
extends to other types of marginal information (see~\cite{JS2} for the basis of
our belief). However, the result about computational efficiency (Theorem
\ref{thm:three}) strongly relies on the efficient description of the first-order
marginal polytope by means of the Birkhoff-Von Neumann theorem and will not readily
extend to other types of marginal information. 
% The result (Theorem \ref{thm:one}) about the universality of the 
% approximation of observed marginal information by sparse choice model relies on 
% somewhat generic proof method and hence will naturally extend beyond the 
% first-order marginal (e.g. for higher-order marginals). In a similar manner, the
% result (Theorem \ref{thm:two}) about approximation by signature family is also likely
% to extend for a large sub-class of choice model (for example, see \cite{JS2} to
% understand why we strongly believe this ought to hold). However, the result about
% computational efficiency  (Theorem \ref{thm:three}) strongly relies on the efficient
% description of the first-order marginal polytope by means of Birkhoff-Von Neumann
% result. 
The algorithm presented in Section~\ref{ssec:algo} extends to higher order
marginals with possibly a computationally complex oracle to check the
feasibility of a signature choice model with respect to the higher order
marginal. Indeed, it would be an important direction for future research to
overcome this computational threshold by possibly developing better
computational approximations. The heuristic utilized in Section 
\ref{sec:empirical} is quite efficient (polynomial in $N$) for first-order
marginals. It is primarily inspired by the exact recovery algorithm based on the signature condition utilized in our earlier work. We strongly believe 
that such a heuristic is likely to provide a computationally efficient procedure
for higher order marginal data.

\subsection{Signature condition and computational efficiency}

As discussed earlier, the signature condition affords us a computationally efficient procedure for the recovery of a sparse choice model approximately consistent with first order marginal information for a broad family of choice models. This is collectively established by Theorems \ref{thm:three}
and \ref{thm:two}. Specifically, the computational speedup relative to brute-force search is significant. However, it is worth asking the 
question whether alternative algorithms that {\em do not} rely on the signature condition can provide a speedup relative to brute-force search. 
%Specifically, can we utilize ideas behind algorithm in Section \ref{ssec:algo} 
%but not use signature condition ?
As it turns out, the following can be shown: Assume there exists a sparse choice model, with sparsity $K$, that approximates the observed first-order marginals (i.e. doubly stochastic matrix) within accuracy $\beps$. In this case, we can recover a sparse choice model (not necessarily in the signature family) with sparsity $O(\beps^{-2} K \log N)$ that approximates the observed first-order marginals within an $\ell_\infty$ error of $O(\beps)$. We can recover this model in time  $\big(\frac{1}{\beps}\big)^K \times \exp\big(K \log K\big)$. Given Theorem \ref{thm:one}
and the following discussion, it is reasonable to think of $K = \Theta(N)$. Therefore,
this computational cost is effectively worse by a factor of $\big(\frac{1}{\beps}\big)^K$ compared to
that of our approach using the signature condition in Section \ref{ssec:algo}. This inferiority notwithstanding, it is worth describing the simple heuristic. Before doing so it is important to note however that this alternate heuristic has no natural generalization to observed data outside of the realm of first-order marginal information. In contrast, the approach we have followed, by {\em relying on the structure afforded by the signature family}, suggests a fast (polynomial in $N$) heuristic that can potentially be applied for many distinct types of marginal data; this is the very heuristic employed in Section \ref{sec:empirical}. Establishing
theoretical guarantees for this heuristic remain an important direction for 
future research. 

%it is important to
%note that this algorithm is almost brute-force and unlike signature condition,
%it does not look for specific structure in the data. That is the reason why it
%may not lead to any better heuristic. On the other hand, as mentioned in
%Section \ref{sec:empirical}, signature condition leads to extremely efficient heuristic
%(polynomial in $N$) which is very promising empirically. Indeed, establishing
%theoretical properties of this heuristic remain an important direction for 
%future research. 

Now we provide a brief description of the algorithm hinted at above. The algorithm is similar to that described in
Section \ref{ssec:algo}. Specifically, it tries to find $K$ permutations
{\em and} their associated probabilities, so that the resulting distribution has first-order
marginals that  are near consistent with the observations. Now the $K$ unknown
permutations are represented through their linear relaxations implied by
the Birkhoff-Von Neumann result, i.e. \eqref{eq:sign1a} and \eqref{eq:sign4a}. 
Under the signature condition, the associated probabilities were discovered 
implicitly by means of \eqref{eq:sign2a} and \eqref{eq:sign3a}. 
However, {\em without} the signature condition, the
only option we have is to search through all possible values for these probabilities. Since
we are interested in approximation accuracy of $\beps$ it suffices to check $\big(K/\beps\big)^K$ such probability vectors. 
%it is sufficient to
%look for all probability values that are quantized into multiple of 
%$\beps/K$. This leads distinct number of $K$ probability values scaling
%as $\big(K/\beps\big)^K$. 
For a given such probability vector, we are left with 
the problem of searching for  $K$ permutations with
these probabilities that have their corresponding first-order marginals
well approximated by the observed data. This again fits into the 
framework of Plotkin, Shmoys and Tardos as discussed in Section 
\ref{ssec:algo}. Therefore, using similar ideas described there, 
we can find a sparse choice model with sparsity $O(\beps^{-2} K \log N)$ 
efficiently (within time polynomial in $N$) if there existed a 
sparse model with sparsity $K$ and the particular quantized probability 
vector that approximated the observations sufficiently well. Since the algorithm will start search over increasing values of 
$K$ and for a given $K$, over all $O((K/\beps)^K)$ distinct probability
vectors, the effective computation cost will be dominated by the 
largest $K$ value encountered by the algorithm. This effectively 
completes the explanation of the algorithm and its computational
cost.

%\author{
\bibliographystyle{plain}
\bibliography{ConcatBib}
%}

\end{document}